\documentclass[12pt,prd,tightenlines,nofootinbib]{revtex4-2}

\usepackage{graphicx}  
\usepackage{dcolumn}   
\usepackage{bm}        
\usepackage{amssymb}   
\usepackage{physics}
\usepackage{makeidx}
\usepackage{dsfont}
\usepackage{relsize}
\usepackage{siunitx}
\usepackage{hyperref}
\usepackage{mathrsfs}
\usepackage{xcolor}
\newcommand{\eq}{\begin{eqnarray}}
\newcommand{\en}{\end{eqnarray}}

\begin{document}

\title{Finite-volume energy shift of the three-nucleon ground state}

\author{Rishabh Bubna}
\affiliation{Helmholtz-Institut f\"ur Strahlen- und Kernphysik (Theorie) and Bethe Center for Theoretical Physics, Universit\"at Bonn, 53115 Bonn, Germany}
\author{Fabian M\"uller}
\affiliation{Helmholtz-Institut f\"ur Strahlen- und Kernphysik (Theorie) and Bethe Center for Theoretical Physics, Universit\"at Bonn, 53115 Bonn, Germany}
 \author{Akaki Rusetsky}
\affiliation{Helmholtz-Institut f\"ur Strahlen- und Kernphysik (Theorie) and Bethe Center for Theoretical Physics, Universit\"at Bonn, 53115 Bonn, Germany}
\affiliation{Tbilisi State  University,  0186 Tbilisi, Georgia\\}


\date{\today}

\begin{abstract}

  A perturbative calculation of the three-nucleon ground-state energy shift in a finite volume
  is carried out within the non-relativistic effective theory. The energy shift is evaluated
  up to and including ${\mathcal O}(L^{-6})$, where $L$ is the size of a cubic box. The convergence
  of the perturbative series at physical values of the scattering lengths is studied numerically.

\end{abstract}

\maketitle

\section{Introduction}

At present, the focus in the lattice studies of the multiparticle systems has been gradually
shifting from two to three particles. In the last few years, three different but equivalent
formalisms known as RFT~\cite{Hansen:2014eka, Hansen:2015zga},
NREFT~\cite{Hammer:2017uqm, Hammer:2017kms} and
FVU~\cite{Mai:2017bge,Mai:2018djl} approaches, have been proposed for the analysis of
data in the three-particle sector. The recent activities in the field~\cite{Kreuzer:2008bi,Kreuzer:2009jp,Kreuzer:2010ti,Kreuzer:2012sr,Briceno:2012rv,Polejaeva:2012ut,Jansen:2015lha,Hansen:2014eka,Hansen:2015zta,Hansen:2015zga,Hansen:2016fzj,Guo:2016fgl,Sharpe:2017jej,Guo:2017crd,Guo:2017ism,Meng:2017jgx,Briceno:2017tce,Hammer:2017uqm,Hammer:2017kms,Mai:2017bge,Guo:2018ibd,Guo:2018xbv,Klos:2018sen,Briceno:2018mlh,Briceno:2018aml,Mai:2019fba,Guo:2019ogp,Guo:2020spn,Blanton:2019igq,Pang:2019dfe,Jackura:2019bmu,Briceno:2019muc,Romero-Lopez:2019qrt,Konig:2020lzo,Brett:2021wyd,Hansen:2020zhy,Blanton:2020gha,Blanton:2020jnm,Pang:2020pkl,Hansen:2020otl,Romero-Lopez:2020rdq,Blanton:2020gmf,Muller:2020vtt,Blanton:2021mih,Muller:2021uur,Beane:2007es,Detmold:2008fn,Detmold:2008yn,Blanton:2019vdk,Horz:2019rrn,Culver:2019vvu,Fischer:2020jzp,Alexandru:2020xqf,Romero-Lopez:2018rcb,Blanton:2021llb,Mai:2021nul,Mai:2018djl,Muller:2022oyw,Blanton:2021eyf,Severt:2022jtg,Baeza-Ballesteros:2023ljl,Draper:2023xvu} include the further development of the formalism as well as actual simulations on the lattice, which are starting to appear.
For more information
on the subject, we refer the reader to the two recent reviews on the
subject~\cite{Hansen:2019nir,Mai:2021lwb}.

Historically, perturbative calculations of the finite-volume shift of the energy levels,
which started in late 1950's~\cite{Huang:1957im,Wu:1959zz,Tan:2007bg,Beane:2007qr,Hansen:2015zta,Hansen:2016fzj,Detmold:2008gh,Muller:2020vtt,Romero-Lopez:2020rdq,Pang:2019dfe}, have predated the derivation of the
exact quantization condition. However, until very recently, all these calculations
were limited to the spin-zero identical particles. Only lately, the perturbative studies were extended to the pions carrying isospin~\cite{Muller:2020vtt}
(the three-body quantization condition for the non-identical particles has been considered
in Refs.~\cite{Blanton:2021eyf,Blanton:2021mih,Blanton:2020gmf}
where, again, the treatment was
restricted to the particles with no spin).

Furthermore, the perturbative expressions for the energy
levels are much easier to use in the fit of the lattice data than the solutions of
the exact quantization condition, and the parameters entering
these expressions can be interpreted in a quite transparent manner. However, the region
of applicability of the perturbative expansion is limited by the ratio of the two-body
scattering length with the box size $L$. This is a
challenge which does not emerge when the exact quantization condition is used.

It is clear that one of the main and potentially very fruitful applications
of the three-body (and many-body) quantization
condition should naturally
be the analysis of the lattice spectra in the three- and many-nucleon
sectors, where a vast abundance of precise experimental data exists. By the same token,
one might argue that the perturbative expansion provides
a tool for the analysis of the energy levels,
which is potentially more simple to handle than the exact quantization condition.
However, the large S-wave $NN$ scattering
lengths pose a very serious obstacle -- the box size $L$, for which the perturbative series
converge, turns out to be unreasonably high, and it is not conceivable that we will be able
to carry out lattice calculations for such large boxes in a foreseeable future.
One could however argue that it might be
still interesting to work out a perturbative expansion, for the following
reasons:

\begin{itemize}

\item
  An intriguing possibility that the light quark masses in Nature
  are close to the critical values, for
  which both singlet and triplet S-wave $NN$ scattering lengths tend to infinity,
  is discussed in the literature (see, e.g., Ref.~\cite{Epelbaum:2008ga}). Turning the
  argument, away from the physical values of quark masses the scattering lengths may
  become of natural size and the lower bound on the box size could get less restrictive.

\item
\begin{sloppypar}
  The available physical volumes in the Nuclear Lattice Effective Field Theory
  (NLEFT) \cite{Lahde:2019npb} at present are much larger than in lattice QCD calculations. This means that the obtained perturbative expressions could be useful in NLEFT, especially if (see the preceding argument) unphysical pion masses are considered.
\end{sloppypar}

\item
  Even the parameter(s) that control 
the convergence of the perturbative expansion in the
two and many-particle sectors are the same, it is still interesting to numerically compare
the convergence speed in the different (two- and three-particle) sectors.

\item
  Conceptually, the presence of an additional scale in the low-energy $NN$ scattering that
  manifests itself in the unnaturally large scattering lengths leads to problems when
  standard dimensional regularization and the minimal subtraction are used. In
  infinite volume, the problem is cured by use of the modified renormalization
  prescriptions, say the PDS
  scheme~\cite{Kaplan:1996xu,Kaplan:1998tg,Kaplan:1998we} or
  the momentum-space subtraction~\cite{Gegelia:1998gn}. The implications of such
  a modification for the finite-volume calculations need to be considered.

\item
  Recently, a full-fledged
  three-particle quantization condition for spin-$1/2$ particles
has become available~\cite{Draper:2023xvu}.
  Hence, even if it turns out that the
  perturbative expression for the ground-state energy shift, which will be derived
  in the present
  paper, cannot be used for the analysis of lattice data, it still
  could serve as a nice testing ground for
the exact
quantization condition after performing the threshold expansion
along the lines of Refs.~\cite{Hansen:2016fzj,Pang:2019dfe}.

\end{itemize}

To summarize, we believe that the arguments given above suffice to
justify the effort invested
in the calculations. We present these calculations in a rather condensed form, since
the method used here is pretty standard by now. The layout of the paper is the following.
In Sect.~\ref{sec:matching} we consider the non-relativistic three-nucleon Lagrangian
and the perturbative matching to the three-particle amplitude. The details of the calculations
of the energy shift are given in Sect.~\ref{sec:energyshift}. Finally,
Sect.~\ref{sec:final} contains our final result and the discussion of the
convergence of the perturbative expansion.

\section{The Lagrangian in NREFT and matching}
\label{sec:matching}
In the calculations, we closely follow the procedure laid out in the
Refs.~\cite{Romero-Lopez:2020rdq,Beane:2007qr,Muller:2020vtt}.
The non-relativistic effective Lagrangian which suffices to carry out calculations
up to and including order $L^{-6}$
is taken from the Refs.~\cite{Chen:1999tn,Bedaque:1999ve} and is given by
\begin{align}\label{eq:L_full}
    &\mathscr{L} \, = \, N^{\dagger} \Big( i\partial_{0}\, + \, \frac{\nabla^2}{2M}\, +\, \frac{\nabla^4}{8M^3}\Big)N\, + \, \mathscr{L}_{2}\,+\, \mathscr{L}_{3}.
\end{align}
Here, the non-relativistic nucleon field $N$ is a doublet in spin as well as in isospin space,
$M$ is the mass of the nucleon and $\mathscr{L}_{2}$ and $\mathscr{L}_{3}$
are the two-nucleon and three-nucleon Lagrangians, respectively. Note that
the isospin and spin indices of the fields have been suppressed for notational
convenience. The relativistic effects are included perturbatively and, at the order we
are working, all these effects boil down to the correction in the kinetic term.

The two-nucleon Lagrangian in Eq.~(\ref{eq:L_full}) is given by\footnote{The Lagrangian that describes P-wave scattering is formally of the same order as the derivative term in the S-wave, which is included here. However, in difference to the latter, the matrix elements of the P-wave Lagrangian
between the two-particle states vanish when at least one of the particle pairs has zero relative momentum. For this reason, the P-waves do not contribute to the ground-state energy shift at the accuracy we are working and are therefore omitted in Eq.~(\ref{eq:L2}).}
\begin{align}\label{eq:L2}
    \mathscr{L}_{2} \, = \, &- C_{0}^{t} (N^{T}R^{t}_{i}N)^{\dagger}(N^{T}R^{t}_{i}N)\,- \, C_{0}^{s}(N^{T}R^{s}_{a}N)^{\dagger}(N^{T}R^{s}_{a}N)\,\nonumber\\[2mm]
     \,+\,& \frac{C^{t}_{2}}{4}\,\Big[ (N^{T}R^{t}_{i}N)^{\dagger} ( N^{T} R^{t}_{i}\overleftrightarrow{\nabla}^{2}N) \,+\, \text{h.c.}\Big]\nonumber\\[2mm]
    \,+\,& \frac{C^{s}_{2}}{4}\,\Big[ (N^{T}R^{s}_{a}N)^{\dagger} ( N^{T}R^{s}_{a}\overleftrightarrow{\nabla}^{2}N)\,+\, \text{h.c.}\Big]\,.
\end{align}
Here, the Galilean-invariant derivative is defined as $\overleftrightarrow{\nabla}\,=\, (\overrightarrow{\nabla} - \overleftarrow{\nabla})/2$. The quantities $R^{t}_{i}$ and $R^{s}_{a}$ are defined as follows (the superscripts $t$ and $s$ stand for the ``triplet'' and ``singlet'' channels, respectively) 
\begin{align}
    R^{t}_{i} \, = \, \sigma_{i}\sigma_{2}\tau_{2}\,,\quad\quad
                   R^{s}_{a}\, = \, \tau_{a}\tau_{2}\sigma_{2}\,.
\end{align}
Here, $\sigma_i$ and $\tau_a$ are the Pauli matrices operating in the spin and
isospin spaces, respectively.
The matrices $R$ are normalized, according to
\begin{align}
\text{Tr}\,\bigl[(R^{t}_{i})^{\dagger}R^{t}_{j}\bigr]\, =\, 4\delta_{ij}\,,\quad\quad
\text{Tr}\,\bigl[(R^{s}_{a})^{\dagger}R^{s}_{b}\bigr]\, =\, 4\delta_{ab}\,,\quad\quad
\text{Tr}\,\bigl[(R^{t}_{i})^{\dagger}R^{s}_{a}\bigr]\, =\, 0\, .
\end{align}
The quantities $C^{t}_{0}$ and $C^{s}_{0}$ are the two-nucleon non-derivative
coupling constants in the $^{3}S_{1}$ and $^{1}S_{0}$ channels respectively. These
constants are matched to parameters of the effective range expansion in the
S-wave $NN$ scattering, which is given by
\begin{align}
  p\cot{\delta_{s,t}(p)}\, =\, -\frac{1}{a_{s,t}} \, +\, \frac{1}{2}r_{s,t}p^{2}\,+\, \dotsc\,.
\end{align}
Here, $a_{s,t}$ represents the $NN$ scattering length and $r_{s,t}$ represents the effective range in the pertinent channel.
The matching condition in the two-body sector takes the form
\begin{align}
    C_{0}^{s,t}\, =\, \frac{\pi}{2M}a_{s,t}\,,\quad\quad
  C_{2}^{s,t}\, =\, \frac{\pi}{2M}a^{2}_{s,t}\Hat{r}_{s,t}\,.
\end{align}
Here, to ease notations, we have defined  $\Hat{r}_{s,t}\, =\, r_{s,t}\, -\,
1/(a_{s,t}M^{2})$. The second term in this expression represents the relativistic
correction~\cite{Romero-Lopez:2020rdq}. We would like to stress here that we use
standard dimensional regularization with minimal subtraction, as if the
scattering lengths were of natural size. Strictly speaking, this constitutes an additional
assumption with respect to $a_{s,t}/L\ll 1$, which guarantees the convergence of the
perturbative expansion. In order to see this, one could imagine for a moment that lattice
simulations are done for physical masses but at very large values of $L$. Then, the
perturbative series will converge but, still, the use of a modified renormalization scheme
(say, the PDS or the momentum-space subtraction scheme)
and a partial resummation of the leading term in the amplitude will be necessary. So,
here one encounters an interesting conceptual problem that one needs to address.

In order to solve this problem, let us first note that the choice of the renormalization
prescription refers to the region of large momenta, whereas the finite-volume effects
emerge from the infrared region. Thus, a clear scale separation emerges and it is
natural to think that these two issues should not be intertwined with each other.
To see that this is indeed the case, let us start from the L\"uscher equation in the
two-particle sector~\cite{Beane:2003yx}. It is written down in terms of physical
observables (the scattering phase shift or the effective range expansion parameters)
rather than the effective couplings and does not refer to the renormalization prescription
anymore. Hence, the perturbative expansion that is obtained from this L\"uscher
equation has exactly the same form as the one in case of the scattering lengths of
natural size, and the manner
how the effective range expansion parameters are expressed through the
couplings in the Lagrangian does not affect the final expression. The same pattern is seen
in the system of three identical scalars~\cite{Pang:2019dfe}. Here, the perturbative
energy shift is expressed through the two-body effective range expansion parameters
and the particle-dimer coupling constant $H_0(\Lambda)$, which can be unambiguously related to the three-particle scattering amplitude at threshold. Hence, once the energy shift
is expressed through the observable $S$-matrix elements, all differences related to the choice of the renormalization prescription and the size of the two-body scattering length are washed out (here, it is assumed that the ratio $a/L$ is small). One could thus expect the same behavior in the three-nucleon case as well.  

After this discussion, let us turn to the matching in the three-particle sector which
becomes relevant at order $L^{-6}$.
The symmetries of the theory allow for three different operators without
derivatives~\cite{Bedaque:1999ve}. 
Using Fierz rearrangement, it can be shown that all the three operators can be reduced to one which can be chosen as follows
\begin{align}
    \mathscr{L}_{3}\, &=\,- \eta_{3}(N^{T}\tau_{2}\sigma_{k}\sigma_{2}N)^{\dagger}(N^{\dagger}\sigma_{k}\sigma_{l}N)(N^{T}\tau_{2}\sigma_{l}\sigma_{2}N)\nonumber\\[2mm]
    &= \, -\eta_{3}(N^{T}R^{t}_{k}N)^{\dagger}(N^{\dagger}\sigma_{k}\sigma_{l}N)(N^{T}R^{t}_{l}N)\,.
\end{align}
The coupling $\eta_{3}$ is ultraviolet-divergent. This divergence is canceled by the divergences that arise in the Feynman diagrams at higher orders in perturbation theory.
The finite part of the coupling $\eta_3$ can be matched to the three-particle threshold amplitude in perturbation theory. 
We use the definition of the threshold amplitude from
Ref.~\cite{Romero-Lopez:2020rdq}. The matching is carried out by considering
the relativistic on-shell amplitude $\mathscr{M}_{3}$ for the process
$p \uparrow p \downarrow n \uparrow \to p \uparrow p \downarrow n \uparrow$, where
$p$, $n$ denote the proton and the neutron,
and the arrows $\uparrow$, $\downarrow$ are the projection of spin on the third axis.

The amplitude $\mathscr{M}_{3}$ can be parameterized in terms of a single variable
$\lambda$ by making a particular choice of incoming and outgoing three-momenta as in Ref.~\cite{Romero-Lopez:2020rdq}:
\begin{align}
    &\textbf{p}_{1}\, = \, \lambda\textbf{e}_{y}\,,\nonumber\\
    &\textbf{p}_{2}\, = \, \lambda\Big(\frac{\sqrt{3}}{2}\textbf{e}_{x}\, -\, \frac{1}{2}\textbf{e}_{y}\Big)\,,\nonumber\\
    &\textbf{p}_{3}\, = \, -\lambda\Big(\frac{\sqrt{3}}{2}\textbf{e}_{x}\, +\, \frac{1}{2}\textbf{e}_{y}\Big)\,,
\end{align}
with $\textbf{p}_{i}\, =\, -\, \textbf{p}'_{i}$ for $i\, =\, 1,2,3$ and $\textbf{e}_{x,y,z}$ representing the appropriate unit vectors in momentum space along the direction of the axes.

In the limit $\lambda\to 0$, this amplitude becomes singular. We can obtain the regular quantity from this amplitude by subtracting the singular terms in a well-defined manner~\cite{Romero-Lopez:2020rdq}:
\begin{align}
   \text{Re}\Big( \mathscr{M}_{3}(\lambda) \, -\, \mathscr{M}_{3}^{(\text{pole})}(\lambda) \Big)\, =\, \frac{1}{\lambda}\mathscr{M}_{3}^{(-1)}\, + \ln{\frac{\lambda}{M}}\mathscr{M}_{3}^{(l)} \, +\, \mathscr{M}_{3}^{(0)}\, + \, \dotsc\,,
\end{align}
where the terms that vanish as $\lambda\to 0$ are represented by the ellipses and
$\mathscr{M}_{3}^{(0)}$ denotes the threshold amplitude. The scale in the logarithm is chosen to be $M$ to simplify the calculation.

The matching procedure is elaborately explained in Ref. \cite{Romero-Lopez:2020rdq}.
Here, we only quote the final result for the finite part of the coupling constant
$\eta_3^r(\mu)$ (the scale of the dimensional regularization $\mu$
could be chosen equal to $M$ for convenience). The relation between $\eta_3$ and $\eta_3^r(\mu)$ is given by
\begin{align}
    \eta_{3}\, = \, &\biggl(\frac{3\sqrt{3}\pi\big\{ a_{s}^{4}\, +\, 6a_{s}^{3}a_{t} \, +\, 18 a_{s}^{2}a_{t}^{2}\, + \, 6a_{s}a_{t}^{3}\, +\, a_{t}^{4} \big\}}{4M}\nonumber\\[2mm]
                    &-\,\frac{\pi^{2}\big\{ a_{s}^{4}\, +\, 24a_{s}^{3}a_{t} \, +\, 78 a_{s}^{2}a_{t}^{2}\, + \, 24a_{s}a_{t}^{3}\, +\, a_{t}^{4} \big\}}{4M}\biggr)
                      \frac{(\mu^{2})^{d-3}}{d \, -\, 3 }
+ \, \eta_{3}^{r}(\mu)\,.
\end{align}
Here, $d$ denotes the number of spatial dimensions.

The matching for $\eta_3^r(\mu)$ gives:
\begin{align}
-  4\eta_{3}^{r}(\mu)\, =\,&
  \frac{1}{(2M)^{3}}\mathscr{M}_{3}^{(0)}\nonumber\\[2mm]
  \,-\,&\frac{6\pi^{2}\big\{ a_{s}^{2}(a_{s} + 3a_{t})\Hat{r}_{s} \, +\, a_{t}^{2}(a_{t} + 3a_{s})\Hat{r}_{t} \big\}}{M}\,-\, \frac{12 \pi^{2}\big\{ a_{s}^{2} \, + \, 6a_{s}a_{t} \, + \, a_{t}^{2} \big\}}{M^{3}}\, +\,
\nonumber\\[2mm]
    \,-\,&\frac{3\sqrt{3}\pi\big\{ a_{s}^{4}\, +\, 6a_{s}^{3}a_{t} \, +\, 18 a_{s}^{2}a_{t}^{2}\, + \, 6a_{s}a_{t}^{3}\, +\, a_{t}^{4} \big\}}{M}\bar\delta_{d}(\mu)\nonumber\\[2mm]
    \,+\,& \frac{\pi^{2}\big\{ a_{s}^{4}\, +\, 24a_{s}^{3}a_{t} \, +\, 78 a_{s}^{2}a_{t}^{2}\, + \, 24a_{s}a_{t}^{3}\, +\, a_{t}^{4} \big\}}{M}\bar\delta_{e}(\mu)\,.\label{eq:1}
\end{align}
Here,
\eq
\bar\delta_{d,e}(\mu)=\delta_{d,e}-\Gamma'(1)-\ln 4\pi+\ln\frac{M^2}{\mu^2}\, ,
\en
and
$\delta_{d} = -1.08964$, $\delta_{e} = 3.92587$ are numerical constants
which can be found in Appendix A of Ref.~\cite{Romero-Lopez:2020rdq}. Thus,
Eq.~(\ref{eq:1}) relates (in perturbation theory)
the renormalized three-nucleon non-derivative coupling
to the regular part of the three-nucleon threshold scattering amplitude, which can
be expressed through the $S$-matrix element and is thus an observable quantity.

\section{The matrix element of the interaction potential in a finite volume}
\label{sec:energyshift}

In a finite volume box of size $L$, the momenta are discretized, $\textbf{p} \, = \, \frac{2\pi}{L}\textbf{n}\,$, where $\textbf{n} \in \mathds{Z}^{3}$.
The free fields in a finite volume can be expanded in Fourier series of the annihilation and creation operators. Namely,
\begin{align}
    &N^h_\alpha (\textbf{x},t)\, = \, \frac{1}{L^{3}}\sum_{\textbf{p}}e^{-ip^{0}t \, +\, i\textbf{px}} a^h_\alpha (\textbf{p})\,,\nonumber\\
    &N^{h\, \dagger}_\alpha (\textbf{x},t)\, = \, \frac{1}{L^{3}}\sum_{\textbf{p}}e^{ip^{0}t \, -\, i\textbf{px}} a^{h\, \dagger}_\alpha (\textbf{p})\,,
\end{align}
where $h=p,n$ and $\alpha=\uparrow,\downarrow$
are the isospin and the spin indices respectively.
The annihilation/creation operators $a^h_\alpha(\textbf{p})\,$,
$a^{h\,\dagger}_\alpha(\textbf{p})\,$
obey the following anticommutation relations:
\begin{align}
  \{a^h_\alpha(\textbf{p}), a^{h'\,\dagger}_{\alpha'}(\textbf{q})\}\,
  = \, L^{3}\delta_{\textbf{pq}}\delta_{hh'}\delta_{\alpha\alpha'}\,.
\end{align}
Before we dive into the perturbative corrections to the ground state energy, we need to define the states that are allowed in a finite volume by Pauli's exclusion principle.
The nucleons are $SU(2)$ doublets in both spin and isospin space. Hence,
three particle states transform in the following way
\begin{align}
    \bm{2}\,\otimes\, \bm{2} \, \otimes\, \bm{2}\, = \, (\bm{1}_{\text{A}}\, \oplus\, \bm{3}_{S}) \, \otimes\, \bm{2}\,=\, \bm{2}_{\text{MA}} \, \oplus\, \bm{2}_{\text{MS}}\, \oplus\, \bm{4}_{S}\,.
\end{align}
Here, A,S denote totally antisymmetric and symmetric states, respectively,
and MA,MS are states with mixed symmetry (antisymmetric/symmetric
with respect to the exchange of the first two particles). 
The complete three-nucleon state should be totally anti-symmetric.
At threshold,  $\textbf{p}_{1,2,3}\,= \, 0\,$. The spin-isospin states should combine,
to give a completely anti-symmetric state. This can be achieved by combining
the spin and isospin states in the following way:
\begin{align}
  \frac{1}{\sqrt{2}} \Big\{(\bm{2}^{I}_{\text{MA}}\, \otimes \, \bm{2}^{S}_{\text{MS}})\, -\,  (\bm{2}^{I}_{\text{MS}}\, \otimes \, \bm{2}^{S}_{\text{MA}})\Big\}\,,
\end{align}
where the superscript $I,S$ stands for the isospin and the spin, respectively.
This combination of representations contains 4 states, corresponding to different
projections of spin $S_3=\pm 1/2$ and isospin $I_3=\pm 1/2$.
An important point to note here
is that these states are orthonormal to each other and do not mix at any order in
perturbation theory, because the Lagrangian has exact spin-isospin symmetry.
Therefore, one needs to perform the calculation, taking any one of these states.
We choose the state with the third component of spin $S_3=+1/2\,$ and isospin $I_3=+1/2\,$.
Then, this state is given by
\begin{align}
    |\text{A} \rangle \, =\, \frac{1}{\sqrt{6}}\Big\{ &| p\uparrow\rangle| n\uparrow\rangle| p\downarrow\rangle \, -\, | p\downarrow\rangle| n\uparrow\rangle| p\uparrow\rangle\, -\, | n\uparrow\rangle| p\uparrow\rangle| p\downarrow\rangle \,\nonumber\\[2mm]
    +\, &| n\uparrow\rangle| p\downarrow\rangle| p\uparrow\rangle\, -\, | p\uparrow\rangle| p\downarrow\rangle| n\uparrow\rangle\,+\, | p\downarrow\rangle| p\uparrow\rangle| n\uparrow\rangle \Big\}\,.\label{state:A}
\end{align}
This is the only possible state with all three-momenta vanishing. There can be three
other states containing the nucleons with nonzero momenta. All these contribute
in the sums over intermediate states and have mixed symmetry with respect to the permutations. Explicit expressions in terms of the single-nucleon wave functions are given below:   
\begin{align}
    | \text{S}\rangle \, =\,\frac{1}{\sqrt{6}}\Big\{ &| p\uparrow\rangle| n\uparrow\rangle| p\downarrow\rangle \, -\, | p\uparrow\rangle| n\downarrow\rangle| p\uparrow\rangle\, -\, | n\uparrow\rangle| p\uparrow\rangle| p\downarrow\rangle \,\nonumber\\[2mm]
    +\, &| n\downarrow\rangle| p\uparrow\rangle| p\uparrow\rangle\, +\, | p\uparrow\rangle| p\downarrow\rangle| n\uparrow\rangle\,-\, | p\downarrow\rangle| p\uparrow\rangle| n\uparrow\rangle \Big\}\,,\nonumber\\[2mm]
     | \text{SS} \rangle \, =\, \frac{1}{6}\Big\{4&| p\uparrow\rangle| p\uparrow\rangle| n\downarrow\rangle\, - \, 2| p\uparrow\rangle| p\downarrow\rangle| n\uparrow\rangle\,-\, 2| p\downarrow\rangle| p\uparrow\rangle| n\uparrow\rangle\,\nonumber\\[2mm]
    -2&| p\uparrow\rangle| n\uparrow\rangle| p\downarrow\rangle\,+\,| p\uparrow\rangle| n\downarrow\rangle| p\uparrow\rangle\,+\, | p\downarrow\rangle| n\uparrow\rangle| p\uparrow\rangle\,\nonumber\\[2mm]
    -2&| n\uparrow\rangle| p\uparrow\rangle| p\downarrow\rangle\,+\, | n\uparrow\rangle| p\downarrow\rangle| p\uparrow\rangle\,+\, | n\downarrow\rangle| p\uparrow\rangle| p\uparrow\rangle\,\Big\}\,,\nonumber\\[2mm]
    |\text{AA}\rangle \, =\, \frac{1}{2}\Big\{ &| p\uparrow\rangle| n\downarrow\rangle| p\uparrow\rangle\, -\, | p\downarrow\rangle| n\uparrow\rangle| p\uparrow\rangle\nonumber\\
    -\,&| n\uparrow\rangle| p\downarrow\rangle| p\uparrow\rangle\, +\, | n\downarrow\rangle| p\uparrow\rangle| p\uparrow\rangle\Big\}\,.\label{state:rest}
\end{align}
Note: all states above have the third component of spin $S_3=+1/2\,$ and isospin $I_3=+1/2\,$. We observe that only states with $I\, =\, 1/2,\,I_{3}\,=\, \pm1/2,\,S\,=\,1/2,\,S_{3}\,=\,\pm1/2\,$ are allowed by Pauli's exclusion principle and the conservation of isospin.
The Lagrangian of NREFT considered here obeys exact spin-isospin symmetry. Consequently,
the states with $I/S\, =\, 3/2\,$ will not couple to the states with $I/S\, = \, 1/2\,$ and hence do not contribute at threshold or as the intermediate states in perturbation theory (even at nonzero momenta).

In order to proceed with perturbative calculations, one needs to construct the
Hamiltonian from the given Lagrangian of the NREFT. This can be easily achieved by use of the canonical procedure and leads to the following result:
\begin{align}
    &\textbf{H}\, =\, \textbf{H}_{0}\,+\, \textbf{H}_{1}\, +\, \textbf{H}_{2}\, +\, \textbf{H}_{3}\, +\, \textbf{H}_{\text{r}}\, =\, \textbf{H}_{0}\,+\,  \textbf{H}_{I}\,
\end{align}
In the above equation \textbf{H}$_{0}\,$ represents the Hamiltonian in the free theory and all the other terms are treated as a perturbation. Individual terms in the above equation
are defined as follows:
\begin{align}
    \textbf{H}_{0}\, &= \, -\int_{L} d^{3}\textbf{x}\,\frac{1}{2M}N^{\dagger}\nabla^{2}N\,,\\[2mm]
    \textbf{H}_{1}\, &= \, \int_{L} d^{3}\textbf{x}\,\Big[C_{0}^{t} (N^{T}R^{t}_{i}N)^{\dagger}(N^{T}R^{t}_{i}N)\,+ \, C_{0}^{s}(N^{T}R^{s}_{a}N)^{\dagger}(N^{T}R^{s}_{a}N)\Big]\,,\\[2mm]
    \textbf{H}_{2}\, &= \,-\int_{L} d^{3}\textbf{x}\,C^{t}_{2}\frac{1}{4}\Big[ (N^{T}R^{t}_{i}N)^{\dagger} ( N^{T} R^{t}_{i}\overleftrightarrow{\nabla}^{2}N) \,+\, h.c.\Big]\nonumber\\[2mm]
    & -\, \int_{L} d^{3}\textbf{x}\,C^{s}_{2}\frac{1}{4}\Big[ (N^{T}R^{s}_{a}N)^{\dagger}( N^{T} R^{s}_{a}\overleftrightarrow{\nabla}^{2}N)\,+\, h.c.\Big]\,,\\[2mm]
    \textbf{H}_{\text{r}}\, &= \,-\int_{L} d^{3}\textbf{x}\,N^{\dagger}\Big[\frac{\nabla^{4}}{8M^3}\Big]N\,,\\[2mm]
    \textbf{H}_{3}\, &=\,\int_{L} d^{3}\textbf{x}\,\eta_{3}(N^{T}R^{t}_{k}N)^{\dagger}(N^{\dagger}\sigma_{k}\sigma_{l}N)(N^{T}R^{t}_{l}N)\,.
\end{align}
All four three-particle states in a finite cubic box, defined above, are the eigenstates of
$\textbf{H}_{0}\,$ and, hence,
\begin{align}
  \textbf{H}_{0} \big| N^{h_{1}}_{\alpha_{1}}(\textbf{p}_{1})
  N^{h_{2}}_{\alpha_{2}}(\textbf{p}_{2})N^{h_{3}}_{\alpha_{3}}(\textbf{p}_{3})\big\rangle
  \, =\, E_{p}\big| N^{h_{1}}_{\alpha_{1}}(\textbf{p}_{1})
  N^{h_{2}}_{\alpha_{2}}(\textbf{p}_{2})N^{h_{3}}_{\alpha_{3}}(\textbf{p}_{3})\big\rangle\,,
\end{align}
with $E_{p}\, = \, \dfrac{1}{2M}\sum\limits_{i=1}^{3}\textbf{p}_{i}^{2}$, where
$E_{p}$ stands for $E(\textbf{p}_{1},\textbf{p}_{2},\textbf{p}_{3})$. These states have the following normalization
\begin{align}
  \big| N^{h_{1}}_{\alpha_{1}}(\textbf{p}_{1})N^{h_{2}}_{\alpha_{2}}(\textbf{p}_{2})
  N^{h_{3}}_{\alpha_{3}}(\textbf{p}_{3})\big\rangle \, =\, \frac{1}{\sqrt{3!}L^{9/2}}
  a^{h_{1}\,\dagger}_{\alpha_{1}}(\textbf{p}_{1})a^{h_{2}\,\dagger}_{\alpha_{2}}(\textbf{p}_{2})a^{h_{3}\,\dagger}_{\alpha_{3}}(\textbf{p}_{3})\big| 0 \big\rangle\,.
\end{align}
We use Rayleigh-Schr$\ddot{\text{o}}$dinger perturbation theory to calculate
the finite-volume shift of the ground state energy level. Up to and including
order $L^{-6}$, this energy shift is given by the expression
\begin{align}
    \Delta E_{n}\, =\, &V_{nn}\, + \, \sum_{p \neq n} \frac{|V_{np}|^{2}}{E_{n} \, -\, E_{p}} \, +\, \sum_{p,q\neq n}\frac{V_{np}V_{pq}V_{qn}}{(E_{n}\,-\,E_{p})(E_{n}\,-\,E_{q})}\, -\, V_{nn} \sum_{p \neq n} \frac{|V_{np}|^{2}}{(E_{n} \, -\, E_{p})^{2}}\nonumber\\[2mm]
    &+ \, \sum_{p,q,k \neq n}\frac{V_{np}V_{pk}V_{kq}V_{qn}}{(E_{n} \, -\, E_{p})(E_{n} \, -\, E_{k})(E_{n} \, -\, E_{q})} \,-\, \sum_{p,q\neq n}\frac{|V_{np}|^{2}}{E_{n} \, -\, E_{p}}\frac{|V_{nq}|^{2}}{(E_{n} \, -\, E_{q})^{2}}\nonumber\\[2mm]
    &-\, 2V_{nn}\sum_{p,q\neq n}\frac{V_{np}V_{pq}V_{qn}}{(E_{n}\,-\, E_{q})^{2}(E_{n}\,-\, E_{p})}\,
    -\, V_{nn}^{2}\sum_{p\neq n}\frac{|V_{np}|^{2}}{(E_{n}\,-\, E_{p})^{3}}\,.
\end{align}
Here, we use the following notation:
\begin{align}
    V_{pq}\, =\, \langle  p | \textbf{H}_{I} |  q \rangle\,.
\end{align}
In the above expression, $|p\rangle\,$ and $|q\rangle\,$
are shorthand notations for
$|p\rangle = | \textbf{p}_{1},\textbf{p}_{2}, \textbf{p}_{3}\rangle$
and
$|q\rangle = | \textbf{q}_{1},\textbf{q}_{2}, \textbf{q}_{3}\rangle$.

Next, we calculate the matrix elements of the potential between different
states given in Eqs.~(\ref{state:A}) and (\ref{state:rest}). We get:
\begin{align}\label{eq:H1}
  &\langle\text{A}, p| \textbf{H}_{1} | \text{A},q \rangle\, =\,
  \frac{4}{3L^{3}}\big[ C_{0}^{s}\, +\, C_{0}^{t} \big](v_{1} \, +\, v_{2} \, +\, v_{3} \, +\, v^{+}_{1}\,+\, v^{+}_{2}\,+\, v^{+}_{3})\,,\nonumber\\[2mm]
  &\langle\text{A},p| \textbf{H}_{1} | \text{S},q \rangle\, =\,
   \frac{1}{3L^{3}}\big[ C_{0}^{s}\, -\, C_{0}^{t} \big](2v_{1} \, +\, 2v_{2} \, -\, 4v_{3} \,-\,3v^{-}_{1}\, -\,3v^{-}_{2}\,-\, v^{+}_{1}\,-\, v^{+}_{2}\,+\, 2v^{+}_{3})\,,\nonumber\\[2mm]
   &\langle\text{A},p| \textbf{H}_{1} | \text{SS},q \rangle\, =\,
    \frac{1}{\sqrt{6}L^{3}}\big[ C_{0}^{s}\, -\, C_{0}^{t} \big](2v_{1} \, -\, 2v_{2} \, +\,v^{-}_{1}\, -\,v^{-}_{2}\,-\, 2v^{-}_{3}\,-\, v^{+}_{1}\,+\, v^{+}_{2})\,,\nonumber\\[2mm]
    &\langle\text{A},p| \textbf{H}_{1} | \text{AA},q \rangle\, =
    -\,\frac{1}{\sqrt{6}L^{3}}\big[ C_{0}^{s}\, -\, C_{0}^{t} \big](2v_{1} \, -\, 2v_{2} \, +\,v^{-}_{1}\, -\,v^{-}_{2}\,-\, 2v^{-}_{3}\,-\, v^{+}_{1}\,+\, v^{+}_{2})\,,\nonumber\\[2mm]
    &\langle\text{S},p| \textbf{H}_{1} | \text{S},q \rangle\, =\,
     \frac{1}{3L^{3}}\big[ C_{0}^{s}\, +\, C_{0}^{t} \big](v_{1} \, +\, v_{2} \, +\, 4v_{3} \,-\,2v^{+}_{1}\, -\,2v^{+}_{2}\,+\, v^{+}_{3})\,,\nonumber\\[2mm]
     &\langle \text{S},p| \textbf{H}_{1} | \text{SS},q \rangle\, =\,
      \frac{1}{\sqrt{6}L^{3}}\big[ C_{0}^{s}\, +\, C_{0}^{t} \big](v_{1} \, -\, v_{2} \, -\,v^{-}_{1}\, +\,v^{-}_{2}\,-\, v^{-}_{3}\,+\, v^{+}_{1}\,-\, v^{+}_{2})\,,\nonumber\\[2mm]
      &\langle \text{S},p| \textbf{H}_{1} | \text{AA},q \rangle\, =
      -\,\frac{1}{\sqrt{6}L^{3}}\big[ C_{0}^{s}\, +\, C_{0}^{t} \big](v_{1} \, -\, v_{2} \, -\,v^{-}_{1}\, +\,v^{-}_{2}\,-\, v^{-}_{3}\,+\, v^{+}_{1}\,-\, v^{+}_{2})\,,\nonumber\\[2mm]
      &\langle \text{SS},p| \textbf{H}_{1} | \text{SS},q \rangle\, =\,
       \frac{1}{2L^{3}}\big[ C_{0}^{s}\, +\, C_{0}^{t} \big](v_{1} \, +\, v_{2} \, -\,v^{+}_{3})\,,\nonumber\\[2mm]
       &\langle \text{SS},p| \textbf{H}_{1} | \text{AA},q \rangle\, =\,
       -\frac{1}{2L^{3}}\big[ C_{0}^{s}\, +\, C_{0}^{t} \big](v_{1} \, +\, v_{2} \, -\,v^{+}_{3})\,,\nonumber\\[2mm]
       &\langle\text{AA},p| \textbf{H}_{1} | \text{AA},q \rangle\, =\,
        \frac{1}{2L^{3}}\big[ C_{0}^{s}\, +\, C_{0}^{t} \big](v_{1} \, +\, v_{2} \, -\,v^{+}_{3})\,,
\end{align}
where
\begin{align}
    &v_{1}\,=\, \delta_{\textbf{p}_{1},\textbf{q}_{1}}\delta_{\textbf{p}_{2}+\textbf{p}_{3},\textbf{q}_{2}+\textbf{q}_{3}}\,,\\[2mm]
    &v_{2}\,=\, \delta_{\textbf{p}_{2},\textbf{q}_{2}}\delta_{\textbf{p}_{1}+\textbf{p}_{3},\textbf{q}_{1}+\textbf{q}_{3}}\,,\\[2mm]
    &v_{3}\,=\, \delta_{\textbf{p}_{3},\textbf{q}_{3}}\delta_{\textbf{p}_{1}+\textbf{p}_{2},\textbf{q}_{1}+\textbf{q}_{2}}\,,\\[2mm]
    &v^{\pm}_{1}\,=\, \delta_{\textbf{p}_{2},\textbf{q}_{3}}\delta_{\textbf{p}_{1}+\textbf{p}_{3},\textbf{q}_{1}+\textbf{q}_{2}}\,\pm\,\delta_{\textbf{p}_{3},\textbf{q}_{2}}\delta_{\textbf{p}_{1}+\textbf{p}_{2},\textbf{q}_{1}+\textbf{q}_{3}}\,,\\[2mm]
    &v^{\pm}_{2}\,=\, \delta_{\textbf{p}_{1},\textbf{q}_{3}}\delta_{\textbf{p}_{2}+\textbf{p}_{3},\textbf{q}_{1}+\textbf{q}_{2}}\,\pm\,\delta_{\textbf{p}_{3},\textbf{q}_{1}}\delta_{\textbf{p}_{1}+\textbf{p}_{2},\textbf{q}_{2}+\textbf{q}_{3}}\,,\\[2mm]
    &v^{\pm}_{3}\,=\, \delta_{\textbf{p}_{1},\textbf{q}_{2}}\delta_{\textbf{p}_{2}+\textbf{p}_{3},\textbf{q}_{1}+\textbf{q}_{3}}\,\pm\,\delta_{\textbf{p}_{2},\textbf{q}_{1}}\delta_{\textbf{p}_{1}+\textbf{p}_{3},\textbf{q}_{2}+\textbf{q}_{3}}\,.
\end{align}
Furthermore, the matrix elements of the operator $\textbf{H}_{2}$ can be obtained from the matrix elements of $\textbf{H}_{1}$ by making the substitutions
\begin{align}
    a_{s,t}\,\to\,& a_{s,t}^{2}\Hat{r}_{s,t}\,,\nonumber\\[2mm]
    \delta_{\textbf{p}_{m},\textbf{q}_{h}}\delta_{\textbf{p}_{n}+\textbf{p}_{o},\textbf{q}_{r}+\textbf{q}_{l}}\, \to\,& \frac{1}{16}\delta_{\textbf{p}_{m},\textbf{q}_{h}}\delta_{\textbf{p}_{n}+\textbf{p}_{o},\textbf{q}_{r}+\textbf{q}_{l}}[(\textbf{p}_{n}\,-\, \textbf{p}_{o})^{2}\,+ \, (\textbf{q}_{r}\, - \, \textbf{q}_{l})^{2}]\,.
\end{align}
The relativistic corrections are given by the matrix elements of the operator $\textbf{H}_{\text{r}}$. The non-vanishing elements are given by:
\begin{align}\label{eq:Hr}
    &\langle\text{A},p| \textbf{H}_{\text{r}} | \text{A},q \rangle\, =\, -\frac{1}{96M^{3}}\sum_{i = 1}^{3}(\textbf{q}_{i}^{4}\, +\, \textbf{p}_{i}^{4})(v_{123}\, +\, v_{132}\, +\, v_{213}\, +\, v_{231}\,+\, v_{312}\, +\, v_{321})\,,\nonumber\\[2mm]
    &\langle\text{S},p| \textbf{H}_{\text{r}} | \text{S},q \rangle\, =\, -\frac{1}{192M^{3}}\sum_{i = 1}^{3}(\textbf{q}_{i}^{4}\, +\, \textbf{p}_{i}^{4})(2v_{123}\, -\, v_{132}\, +\, 2v_{213}\, -\, v_{231}\,-\, v_{312}\, -\, v_{321})\,,\nonumber\\[2mm]
    &\langle\text{S},p| \textbf{H}_{\text{r}} | \text{SS},q \rangle\, =\, -\frac{1}{64\sqrt{6}M^{3}}\sum_{i = 1}^{3}(\textbf{q}_{i}^{4}\, +\, \textbf{p}_{i}^{4})(v_{132}\, -\, v_{231}\,+\, v_{312}\, -\, v_{321})\,,\nonumber\\[2mm]
    &\langle\text{S},p| \textbf{H}_{\text{r}} | \text{AA},q \rangle\, =\, \frac{1}{64\sqrt{6}M^{3}}\sum_{i = 1}^{3}(\textbf{q}_{i}^{4}\, +\, \textbf{p}_{i}^{4})(v_{132}\, -\, v_{231}\,+\, v_{312}\, -\, v_{321})\,,\nonumber\\[2mm]
    &\langle\text{SS},p| \textbf{H}_{\text{r}} | \text{SS},q \rangle\, =\, -\frac{1}{384M^{3}}\sum_{i = 1}^{3}(\textbf{q}_{i}^{4}\, +\, \textbf{p}_{i}^{4})(4v_{123}\, -\,v_{132}\,-\, 4v_{213}\,+\,v_{231}\,+\, v_{312}\, -\, v_{321})\,,\nonumber\\[2mm]
    &\langle\text{SS},p| \textbf{H}_{\text{r}} | \text{AA},q \rangle\, =\, \frac{1}{128M^{3}}\sum_{i = 1}^{3}(\textbf{q}_{i}^{4}\, +\, \textbf{p}_{i}^{4})(v_{132}\,-\, v_{231}\,-\, v_{312}\, +\, v_{321})\,,\nonumber\\[2mm]
    &\langle\text{AA},p| \textbf{H}_{\text{r}} | \text{AA},q \rangle\, =\, -\frac{1}{384M^{3}}\sum_{i = 1}^{3}(\textbf{q}_{i}^{4}\, +\, \textbf{p}_{i}^{4})(4v_{123}\, -\,v_{132}\,-\, 4v_{213}\,+\,v_{231}\,+\, v_{312}\, -\, v_{321})\,,
\end{align}
where we have
\begin{align}
  v_{ijk}\,=\, \delta_{\textbf{p}_{1},\textbf{q}_{i}}\delta_{\textbf{p}_{2},\textbf{q}_{j}}\delta_{\textbf{p}_{3},\textbf{q}_{k}}\,.
  \end{align}
It is interesting to note also that defining the linear combination
  of two states $|X\rangle=(|SS\rangle-|AA\rangle)/\sqrt{2}$ and
  $|Y\rangle=(|SS\rangle+|AA\rangle)/\sqrt{2}$, the state $|Y\rangle$ completely
  decouples in the matrix elements of ${\bf H}_{1,2}$ (Eq.~(\ref{eq:H1})), but not
  in the matrix elements of ${\bf H}_{\text{r}}$ (Eq.~(\ref{eq:Hr})). Still, at the order we are working,
  the matrix elements of ${\bf H}_{\text{r}}$ containing at least one $|Y\rangle$ state
  do not contribute to the energy shift. Hence, at this order, the whole potential
  effectively reduces to the $3\times 3$ matrix.

Finally, the matrix element for the three body interaction term in the Hamiltonian
is given by
\begin{align}
  \langle\text{A},p| \textbf{H}_{3} | \text{A},q \rangle\, =\,
  \frac{4}{L^{6}}\,\eta_{3}\,\delta_{\textbf{p}_{1}+\textbf{p}_{2}+\textbf{p}_{3},\textbf{q}_{1}+\textbf{q}_{2}+\textbf{q}_{3}}\,.
\end{align}
This is the only non-zero matrix element for the three-body force term that contributes at threshold. 

\section{Perturbative shift to the ground state energy}
\label{sec:final}

We can directly apply the Rayleigh-Schr$\ddot{\text{o}}$dinger perturbation theory to calculate the shift in the ground state energy up to order $L^{-6}$.
To this end, we make use of the fact that $(E\, -\, E_{0})\,$ is of the $\mathcal{O}(L^{-3})\,$. This gives
\begin{align}
    E\, -\, E_{0}\, =\, \frac{E_{3}}{L^{3}}\, +\, \frac{E_{4}}{L^{4}}\, +\, \frac{E_{5}}{L^{5}}\,+\,\frac{E_{6}}{L^{6}}\,+\, \mathcal{O}(L^{-7})\,,
\end{align}
where
\begin{align}
    &\frac{E_{3}}{L^{3}}\, = \, V^{1}_{00}\,,\\[2mm]
    &\frac{E_{4}}{L^{4}}\, = \, -\,\sum_{p \neq 0}\frac{V^{1}_{0p}V^{1}_{p0}}{E_{p}}\,,\label{eq:e4}\\[2mm]
    &\frac{E_{5}}{L^{5}}\, = \, \sum_{p,q\neq 0}\frac{V^{1}_{0p}V^{1}_{pq}V^{1}_{q0}}{E_{p}E_{q}}\,-\,  V^{1}_{00}\sum_{p\neq 0}\frac{V^{1}_{0p}V^{1}_{p0}}{E_{p}^{2}}\,,\\[2mm]
    &\frac{E_{6}}{L^{6}}\, = \, -\,\sum_{p,q,k \neq 0} \frac{V^{1}_{0p}V^{1}_{pk}V^{1}_{kq}V^{1}_{q0}}{E_{p}E_{k}E_{q}}\, -\, \sum_{p \neq 0} \frac{V^{1}_{0p}V^{2}_{p0}\,+\, V^{2}_{0p}V^{1}_{p0}}{E_{p}}\, +\, V^{3}_{00}\, +\, \sum_{p,q \neq 0} \frac{V^{1}_{0p}V^{\text{r}}_{pq}V^{1}_{q0}}{E_{p}E_{q}}\nonumber\\[2mm]
    &\hspace{30pt}\, +\, 2V^{1}_{00}\sum_{p,q\neq 0}\frac{V^{1}_{0p}V^{1}_{pq}V^{1}_{q0}}{E^{2}_{p}E_{q}}\,+\, \sum_{p,q\neq 0}\frac{V^{1}_{0p}V^{1}_{p0}}{E^{2}_{p}}\frac{V^{1}_{0q}V^{1}_{q0}}{E_{q}}\, +\, (V^{1}_{00})^{2}\sum_{p \neq 0}\frac{V^{1}_{0p}V^{1}_{p0}}{E^{3}_{p}}\,.
\end{align}
Here, $V^X_{nm}\,$ represents the matrix element of the operator $\textbf{H}_X$ with
$X=1,2,r,3$. The momentum sums that appear in the above equations are,
in general, ultraviolet divergent. To be consistent with matching,
the divergence has to be tackled with dimensional regularization and
$\overline{\text{MS}}$ renormalization scheme, similar to the infinite-volume case.
The technical details are outlined in Refs.~\cite{Romero-Lopez:2020rdq,Beane:2007qr}.
Plugging in the matrix elements and simplifying, we get the following result
\begin{align}
    &E_{3}\,=\,  \frac{6\pi}{M}(a_{s}\, +\, a_{t})\,,\nonumber\\[2mm]
    &E_{4}\,=\, -\, \frac{6}{M}(a^{2}_{s}\, +\, a^{2}_{t})I\,,\nonumber\\[2mm]
    &E_{5}\,=\, \frac{3}{2M\pi}\Big[ 4(a^{3}_{s}\, +\, a^{3}_{t})I^{2}\, -\, (a_{s}\, + \, a_{t})(5a^{2}_{s}\, -\, 14a_{s}a_{t}\,+\, 5a^{2}_{t})J \Big]\,,\nonumber\\[2mm]
    &E_{6}\,=\,\frac{6}{M\pi^{2}}\Big[ -(a^{4}_{s}\,+\,a^{4}_{t})I^{3}\,+\, (4a^{4}_{s}\,-\, 3a^{3}_{s}a_{t}\,-\, 3a_{s}a^{3}_{t}\,+\, 4a^{4}_{t})IJ\nonumber\\[2mm]
    &\hspace{30pt}-\, \frac{3}{2}(a_{s}\,+\, a_{t})^{2}(a^{2}_{s}\,-\,7a_{s}a_{t}\,+\,a^{2}_{t})K\,-\,\frac{1}{4}(a^{4}_{s}\,+\, 24a^{3}_{s}a_{t}\,+\,78a^{2}_{s}a^{2}_{t}\,+\,24a_{s}a^{3}_{t}\,+\, a^{4}_{t})Q^{r}\nonumber\\[2mm]
    &\hspace{30pt}-\, \frac{1}{2}(a^{4}_{s}\,+\, 6a^{3}_{s}a_{t}\,+\,18a^{2}_{s}a^{2}_{t}\,+\,6a_{s}a^{3}_{t}\,+\, a^{4}_{t})R^{r}\Big]\nonumber\\[2mm]
    &\hspace{30pt}+\, \frac{6\pi^{2}}{M^{3}}(a^{2}_{s}\,+\,a^{2}_{t})\,+\, \frac{12\pi^{2}}{M}(a^{3}_{s}\Hat{r}_{s}\,+\,a^{3}_{t}\Hat{r}_{t})\nonumber\\[2mm]
    &\hspace{30pt}+\, \frac{2\pi}{M}\Big[ 3\sqrt{3}\big\{ a_{s}^{4}\, +\, 6a_{s}^{3}a_{t} \, +\, 18 a_{s}^{2}a_{t}^{2}\, + \, 6a_{s}a_{t}^{3}\, +\, a_{t}^{4} \big\} \nonumber\\[2mm]
    &\hspace{30pt}-\, \pi\big\{ a_{s}^{4}\, +\, 24a_{s}^{3}a_{t} \, +\, 78 a_{s}^{2}a_{t}^{2}\, + \, 24a_{s}a_{t}^{3}\, +\, a_{t}^{4} \big\} \Big]\ln{\big(\mu L\big)} \,+\,4\eta^{r}_{3}(\mu)\,.\label{eqimp}
\end{align}
In the final result, the quantity $\eta^r_{3}(\mu)$ in $E_6$ should be replaced
by $\mathscr{M}_3^{(0)}$ through the matching condition, Eq.~(\ref{eq:1}).
The dependence on $\mu$ disappears, as it should.

In the above expressions, $I,J,K$ are the finite quantities that arise from regularizing the above mentioned momentum sums. These are given by
\begin{align}
    &I \,=\, \sum_{\textbf{n}\neq 0} \frac{1}{\textbf{n}^{2}}\,=\, -8.91363291781\dotsc\,,\nonumber\\[2mm]
    &J \,=\, \sum_{\textbf{n}\neq 0} \frac{1}{\textbf{n}^{4}}\,=\, 16.532315959\dotsc\,,\nonumber\\[2mm]
    &K \,=\, \sum_{\textbf{n}\neq 0} \frac{1}{\textbf{n}^{6}}\,=\, 8.401923974433\dotsc\,.
\end{align}
$Q^{r}\,$ and $R^{r}\,$ are the finite parts obtained after renormalizing the double sums over momenta using the $\overline{\text{MS}}$ scheme. These are given by

\begin{align}
  &\frac{1}{L^{2d}}\sum_{\textbf{p},\textbf{q} \neq 0}\frac{1}{\textbf{p}^{2}\textbf{q}^{2}\big(\textbf{p}^{2}\,+\,\textbf{q}^{2}\,+\,(\textbf{p}\,+\,\textbf{q})^{2}\big)}\nonumber\\[2mm]
  \,=\,& \mu^{2(d\,-\,3)}\Big\{ \frac{1}{48\pi^{2}}\Big( \ln{(\mu L)}\,-\,\frac{1}{2(d\,-\,3)} \Big)\,+\,\frac{1}{(2\pi)^{6}}Q^{r} \Big\}\,,\\[2mm]
  &\frac{1}{L^{2d}}\sum_{\textbf{p} \neq 0}\frac{1}{\textbf{p}^{4}}\sum_{\textbf{q}}\frac{1}{\big(\textbf{p}^{2}\,+\,\textbf{q}^{2}\,+\,(\textbf{p}\,+\,\textbf{q})^{2}\big)}\nonumber\\[2mm]
  \,=\,& \mu^{2(d\,-\,3)}\Big\{ -\frac{\sqrt{3}}{32\pi^{3}}\Big( \ln{(\mu L)}\,-\,\frac{1}{2(d\,-\,3)} \Big)\,+\,\frac{1}{(2\pi)^{6}}R^{r} \Big\}\,.
\end{align}
Numerical values of the finite quantities are given by
\begin{align}
    &Q^{r}\,=\, -102.1556055\dotsc\,,\nonumber\\[2mm]
    &R^{r}\,=\, 19.186903\dotsc\,.
\end{align}
The calculation of these quantities is described in Refs.~\cite{Romero-Lopez:2020rdq,Muller:2020vtt,Beane:2007qr}, where these values are taken from. Note that the
divergent term of the three-body coupling $\eta_3$ is exactly the one that is needed to cancel the divergence in the finite-volume sums. This serves as a nice check of the
calculations. At the end, using Eq.~(\ref{eq:1}), one may replace $\eta_3^r(\mu)$ by the
threshold amplitude $\mathscr{M}_3^{(0)}$, expressing the energy shift solely in terms
of the observable $S$-matrix elements. In this form, the final perturbative
formula is valid in case
of unnaturally large scattering lengths and very large values of $L$.

\begin{figure}
    \centering
   \includegraphics[scale = 0.5]{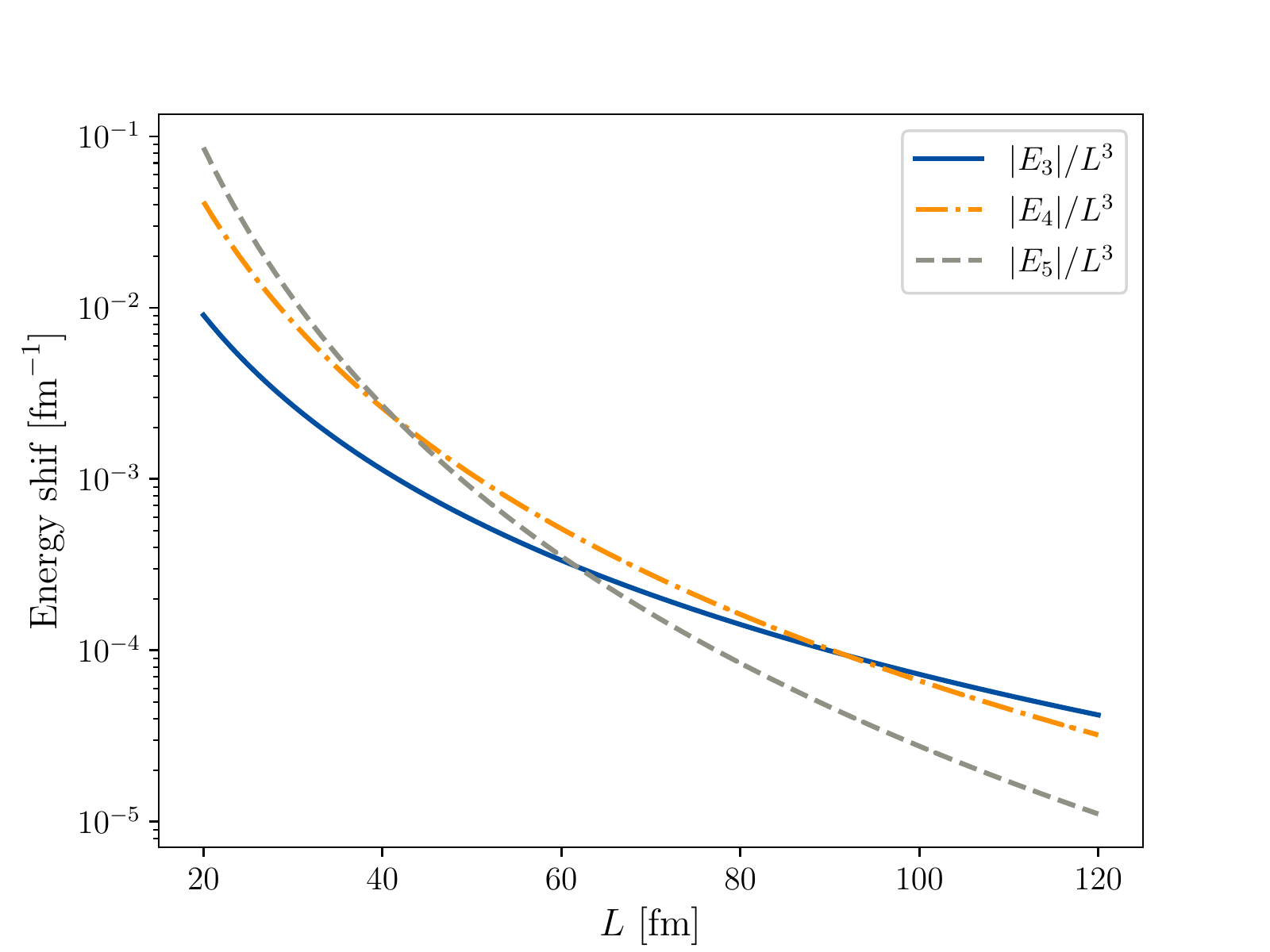}
    \caption{The three-particle energy shift at different orders plotted against the box size $L$.}
    \label{fig:1}
  \end{figure}

\begin{figure}
    \centering
    \includegraphics[scale = 0.5]{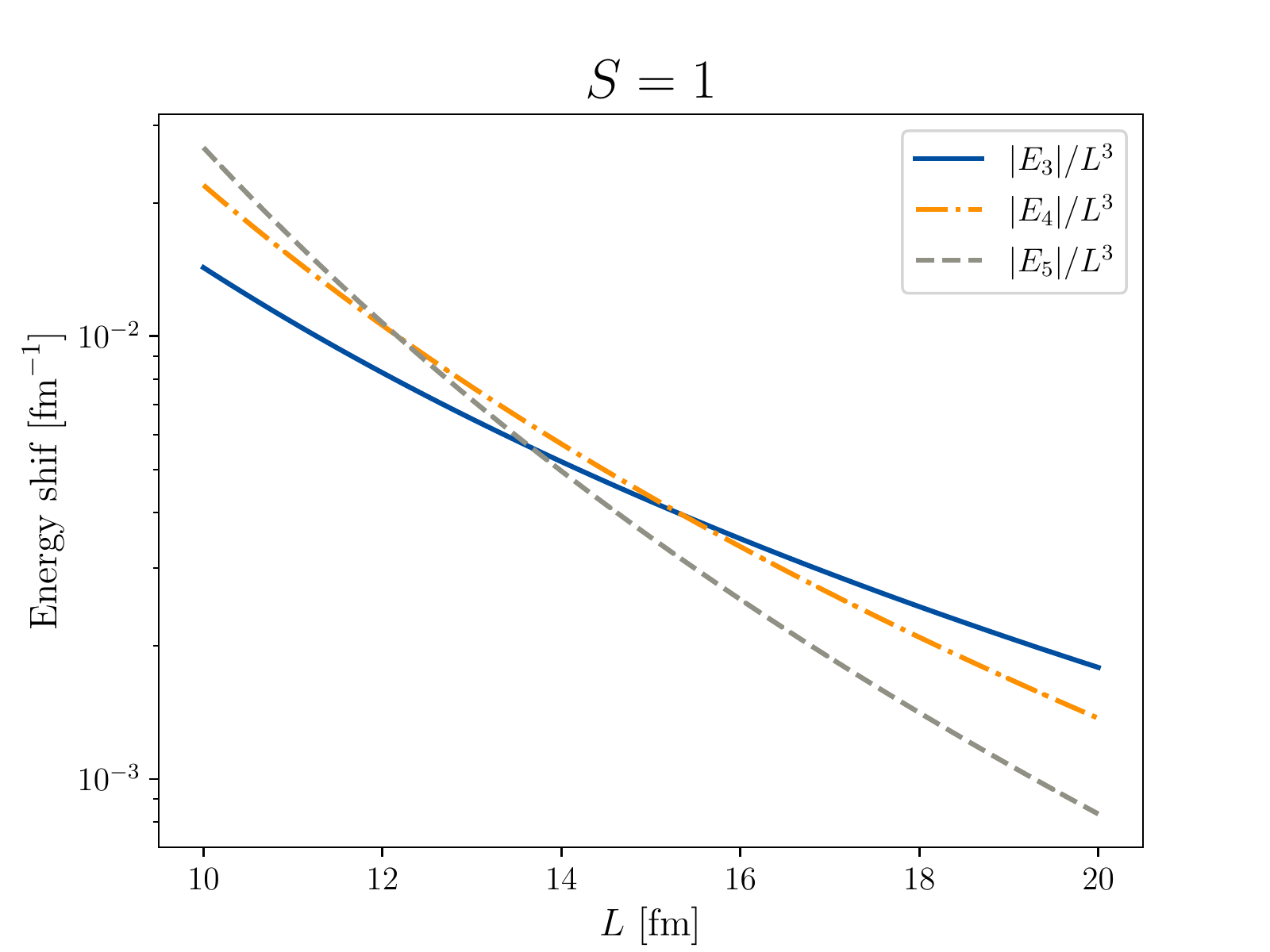}\vspace*{.4cm}
    \includegraphics[scale = 0.5]{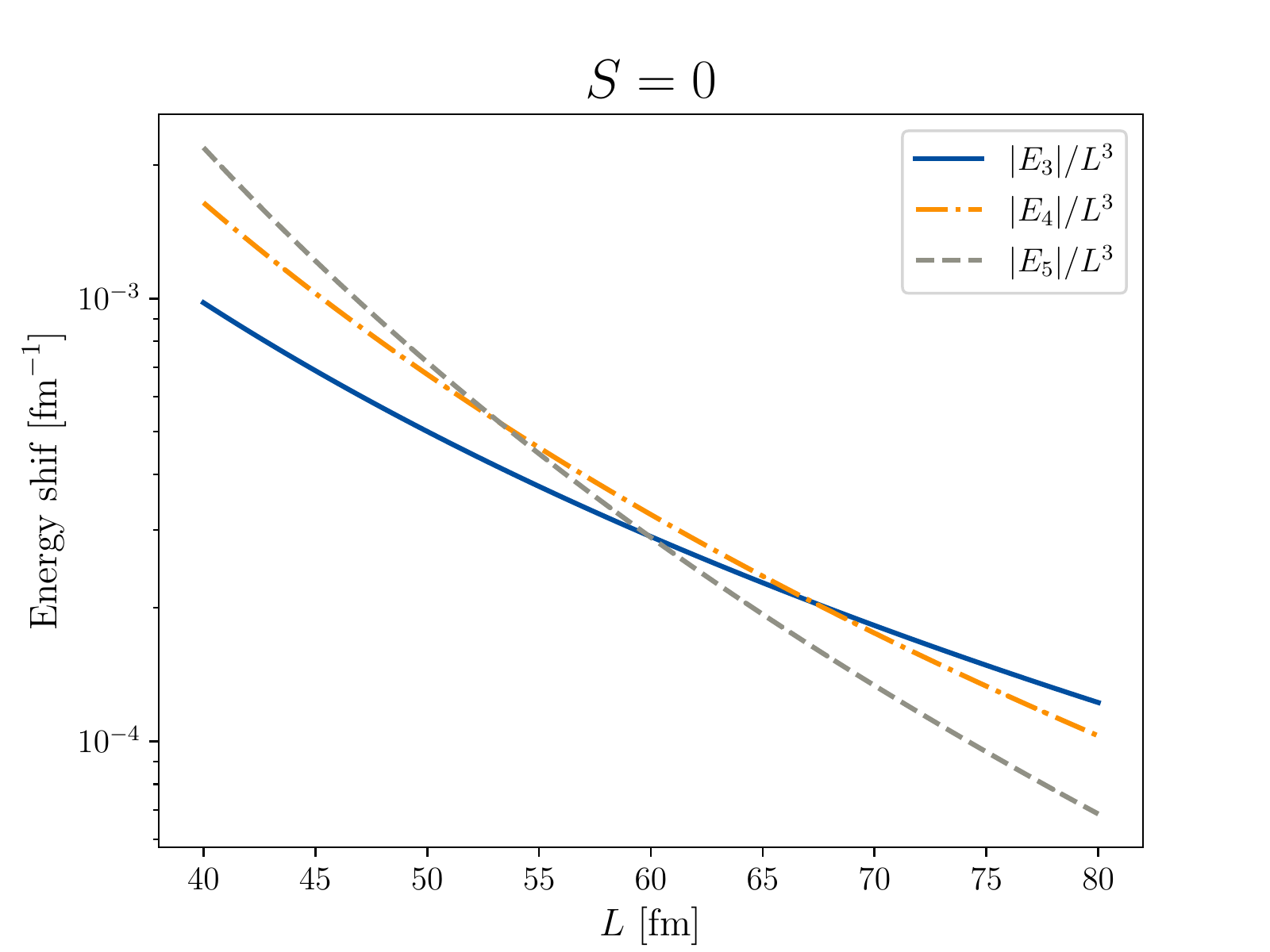}
    \caption{The two-particle energy shift in the spin triplet and singlet channels,
      at different orders, against the box size $L$.}
    \label{fig:2}
  \end{figure}

We find it very instructive to study the convergence of the obtained perturbative formula
for different values of $L$. As an input, we use the following values of the two-body
scattering lengths \cite{Hackenburg:2006qd}:
\begin{align}
    a_{s}\,=\, -23.7148(43) \SI{}{fm}\,,\quad\quad  
    a_{t}\,=\, 5.4112(15) \SI{}{fm}\,, 
\end{align}
In Fig.~\ref{fig:1} we plot $E_{3}/L^{3}, E_{4}/L^{4}$ and $E_{5}/L^{5}$ as
functions of $L$. We do not consider $E_{6}$ in order to avoid
blurring the discussion due to the inclusion of the three-body coupling.
For comparison, in Fig.~\ref{fig:2} we plot the two-particle energy shift in the triplet and singlet channels, using formulae from, e.g., Ref.~\cite{Beane:2007qr}\footnote{In that paper, the expressions for two identical scalars are given.
However, to the order we are working, the nucleon-nucleon
Lippmann-Schwinger equation in the
singlet/triplet channels looks identical to its scalar counterpart. Consequently, one can use the formulae from Ref.~\cite{Beane:2007qr} also here.}.

  From the plots in Fig. \ref{fig:1} it is seen that, for $L \lesssim\, 45\, \SI{}{fm}\,$,
  ${E_{5}}/{L^{5}}\, >\,{E_{4}}/{L^{4}}\,>\,{E_{3}}/{L^{3}}\,$.
  This clearly indicates that, in this case, perturbation theory is not applicable for small $L$.
  For $L$ above $ \SI{90}{fm}\,$ the ordering of the corrections to the ground state
  energy seems to be correct and, naively, one would assume that perturbation theory is applicable above this value. However, the convergence is expected to be very slow
  and hence, in order to achieve a decent convergence rate, one would need to go
  to the very high values of $L$. Furthermore, comparison to Fig.~\ref{fig:2} shows that
  the convergence in the two-particle sector is much faster, albeit still beyond reach for
  the present-day lattice QCD calculations.

\section{Conclusions, outlook}

\begin{itemize}

\item[i)]
  Using the non-relativistic effective theory, 
  we have derived a perturbative expansion of the finite-volume shift of the
  three-nucleon ground-state energy
  up to and including order $L^{-6}$. The matching to the threshold three-nucleon
  amplitude has also been carried out in perturbation theory.

\item[ii)]
  It is known that S-wave nucleon-nucleon scattering lengths have unnaturally large values
  that affects the convergence of the standard perturbative series. As a result,
  the lowest-order term to the two-body amplitude should be resummed to all orders.
  Despite this fact, a strict perturbative derivation of the finite-volume shift is still valid.
  The difference is hidden in expressing of the two and three-body observables
  through the effective couplings and does not affect the final formula of the energy
  shift in terms of the observables.
  
\item[iii)]
  Numerically, at physical values of the scattering lengths, the perturbative series converges
  only at unrealistically high values of $L$. Furthermore, the convergence in the
  three-particle sector is substantially slower than in case of two nucleons.

\item[iv)]
  A full-fledged three-fermion quantization condition has become available recently~\cite{Draper:2023xvu}. In our opinion,
  in the future, the derived perturbative expression may provide a useful testing ground
  for the general framework, controlling, in particular, a consistent inclusion of the
  relativistic effects for particles with spin. These effects are technically challenging in
  the explicitly relativistic-invariant framework and are almost trivial in the non-relativistic
  setting used here.

\end{itemize}

\begin{acknowledgments}
  The authors would like to thank Evan Berkowitz, Serdar Elhatisari,
  Hans-Werner Hammer, Tom Luu
and Steve Sharpe for useful discussions.
This work is supported by the Deutsche Forschungsgemeinschaft (DFG, German Research Foundation) through the funds provided to the Sino-German Collaborative Research Center CRC 110 ``Symmetries and the Emergence of Structure in QCD''
(DFG Project-ID 196253076 -
TRR 110) and by the programme ``Netzwerke 2021'',
an initiative of the Ministry of Culture and Science of the State of North Rhine-Westphalia.
In addition, the work of AR was funded in part by the Volkswagenstiftung (grant no. 93562) and the Chinese Academy of Sciences (CAS) President’s International Fellowship Initiative (PIFI) (grant no. 2021VMB0007).
\end{acknowledgments}

\bibliographystyle{apsrev4-1}
\bibliography{ref1}

\begin{thebibliography}{79}%
\makeatletter
\providecommand \@ifxundefined [1]{%
 \@ifx{#1\undefined}
}%
\providecommand \@ifnum [1]{%
 \ifnum #1\expandafter \@firstoftwo
 \else \expandafter \@secondoftwo
 \fi
}%
\providecommand \@ifx [1]{%
 \ifx #1\expandafter \@firstoftwo
 \else \expandafter \@secondoftwo
 \fi
}%
\providecommand \natexlab [1]{#1}%
\providecommand \enquote  [1]{``#1''}%
\providecommand \bibnamefont  [1]{#1}%
\providecommand \bibfnamefont [1]{#1}%
\providecommand \citenamefont [1]{#1}%
\providecommand \href@noop [0]{\@secondoftwo}%
\providecommand \href [0]{\begingroup \@sanitize@url \@href}%
\providecommand \@href[1]{\@@startlink{#1}\@@href}%
\providecommand \@@href[1]{\endgroup#1\@@endlink}%
\providecommand \@sanitize@url [0]{\catcode `\\12\catcode `\$12\catcode
  `\&12\catcode `\#12\catcode `\^12\catcode `\_12\catcode `\%12\relax}%
\providecommand \@@startlink[1]{}%
\providecommand \@@endlink[0]{}%
\providecommand \url  [0]{\begingroup\@sanitize@url \@url }%
\providecommand \@url [1]{\endgroup\@href {#1}{\urlprefix }}%
\providecommand \urlprefix  [0]{URL }%
\providecommand \Eprint [0]{\href }%
\providecommand \doibase [0]{http://dx.doi.org/}%
\providecommand \selectlanguage [0]{\@gobble}%
\providecommand \bibinfo  [0]{\@secondoftwo}%
\providecommand \bibfield  [0]{\@secondoftwo}%
\providecommand \translation [1]{[#1]}%
\providecommand \BibitemOpen [0]{}%
\providecommand \bibitemStop [0]{}%
\providecommand \bibitemNoStop [0]{.\EOS\space}%
\providecommand \EOS [0]{\spacefactor3000\relax}%
\providecommand \BibitemShut  [1]{\csname bibitem#1\endcsname}%
\let\auto@bib@innerbib\@empty
\bibitem [{\citenamefont {Hansen}\ and\ \citenamefont
  {Sharpe}(2014)}]{Hansen:2014eka}%
  \BibitemOpen
  \bibfield  {author} {\bibinfo {author} {\bibfnamefont {M.~T.}\ \bibnamefont
  {Hansen}}\ and\ \bibinfo {author} {\bibfnamefont {S.~R.}\ \bibnamefont
  {Sharpe}},\ }\href {\doibase 10.1103/PhysRevD.90.116003} {\bibfield
  {journal} {\bibinfo  {journal} {Phys. Rev.}\ }\textbf {\bibinfo {volume}
  {D90}},\ \bibinfo {pages} {116003} (\bibinfo {year} {2014})}\BibitemShut
  {NoStop}%
\bibitem [{\citenamefont {Hansen}\ and\ \citenamefont
  {Sharpe}(2015)}]{Hansen:2015zga}%
  \BibitemOpen
  \bibfield  {author} {\bibinfo {author} {\bibfnamefont {M.~T.}\ \bibnamefont
  {Hansen}}\ and\ \bibinfo {author} {\bibfnamefont {S.~R.}\ \bibnamefont
  {Sharpe}},\ }\href {\doibase 10.1103/PhysRevD.92.114509} {\bibfield
  {journal} {\bibinfo  {journal} {Phys. Rev.}\ }\textbf {\bibinfo {volume}
  {D92}},\ \bibinfo {pages} {114509} (\bibinfo {year} {2015})}\BibitemShut
  {NoStop}%
\bibitem [{\citenamefont {Hammer}\ \emph
  {et~al.}(2017{\natexlab{a}})\citenamefont {Hammer}, \citenamefont {Pang},\
  and\ \citenamefont {Rusetsky}}]{Hammer:2017uqm}%
  \BibitemOpen
  \bibfield  {author} {\bibinfo {author} {\bibfnamefont {H.-W.}\ \bibnamefont
  {Hammer}}, \bibinfo {author} {\bibfnamefont {J.-Y.}\ \bibnamefont {Pang}}, \
  and\ \bibinfo {author} {\bibfnamefont {A.}~\bibnamefont {Rusetsky}},\ }\href
  {\doibase 10.1007/JHEP09(2017)109} {\bibfield  {journal} {\bibinfo  {journal}
  {JHEP}\ }\textbf {\bibinfo {volume} {09}},\ \bibinfo {pages} {109} (\bibinfo
  {year} {2017}{\natexlab{a}})}\BibitemShut {NoStop}%
\bibitem [{\citenamefont {Hammer}\ \emph
  {et~al.}(2017{\natexlab{b}})\citenamefont {Hammer}, \citenamefont {Pang},\
  and\ \citenamefont {Rusetsky}}]{Hammer:2017kms}%
  \BibitemOpen
  \bibfield  {author} {\bibinfo {author} {\bibfnamefont {H.~W.}\ \bibnamefont
  {Hammer}}, \bibinfo {author} {\bibfnamefont {J.~Y.}\ \bibnamefont {Pang}}, \
  and\ \bibinfo {author} {\bibfnamefont {A.}~\bibnamefont {Rusetsky}},\ }\href
  {\doibase 10.1007/JHEP10(2017)115} {\bibfield  {journal} {\bibinfo  {journal}
  {JHEP}\ }\textbf {\bibinfo {volume} {10}},\ \bibinfo {pages} {115} (\bibinfo
  {year} {2017}{\natexlab{b}})}\BibitemShut {NoStop}%
\bibitem [{\citenamefont {Mai}\ and\ \citenamefont
  {{D\"{o}ring}}(2017)}]{Mai:2017bge}%
  \BibitemOpen
  \bibfield  {author} {\bibinfo {author} {\bibfnamefont {M.}~\bibnamefont
  {Mai}}\ and\ \bibinfo {author} {\bibfnamefont {M.}~\bibnamefont
  {{D\"{o}ring}}},\ }\href {\doibase 10.1140/epja/i2017-12440-1} {\bibfield
  {journal} {\bibinfo  {journal} {Eur. Phys. J.}\ }\textbf {\bibinfo {volume}
  {A53}},\ \bibinfo {pages} {240} (\bibinfo {year} {2017})}\BibitemShut
  {NoStop}%
\bibitem [{\citenamefont {Mai}\ and\ \citenamefont
  {D{\"{o}}ring}(2019)}]{Mai:2018djl}%
  \BibitemOpen
  \bibfield  {author} {\bibinfo {author} {\bibfnamefont {M.}~\bibnamefont
  {Mai}}\ and\ \bibinfo {author} {\bibfnamefont {M.}~\bibnamefont
  {D{\"{o}}ring}},\ }\href {\doibase 10.1103/PhysRevLett.122.062503} {\bibfield
   {journal} {\bibinfo  {journal} {Phys. Rev. Lett.}\ }\textbf {\bibinfo
  {volume} {122}},\ \bibinfo {pages} {062503} (\bibinfo {year}
  {2019})}\BibitemShut {NoStop}%
\bibitem [{\citenamefont {Kreuzer}\ and\ \citenamefont
  {Hammer}(2009)}]{Kreuzer:2008bi}%
  \BibitemOpen
  \bibfield  {author} {\bibinfo {author} {\bibfnamefont {S.}~\bibnamefont
  {Kreuzer}}\ and\ \bibinfo {author} {\bibfnamefont {H.~W.}\ \bibnamefont
  {Hammer}},\ }\href {\doibase 10.1016/j.physletb.2009.02.035} {\bibfield
  {journal} {\bibinfo  {journal} {Phys. Lett. B}\ }\textbf {\bibinfo {volume}
  {673}},\ \bibinfo {pages} {260} (\bibinfo {year} {2009})}\BibitemShut
  {NoStop}%
\bibitem [{\citenamefont {Kreuzer}\ and\ \citenamefont
  {Hammer}(2010)}]{Kreuzer:2009jp}%
  \BibitemOpen
  \bibfield  {author} {\bibinfo {author} {\bibfnamefont {S.}~\bibnamefont
  {Kreuzer}}\ and\ \bibinfo {author} {\bibfnamefont {H.~W.}\ \bibnamefont
  {Hammer}},\ }\href {\doibase 10.1140/epja/i2010-10910-6} {\bibfield
  {journal} {\bibinfo  {journal} {Eur. Phys. J. A}\ }\textbf {\bibinfo {volume}
  {43}},\ \bibinfo {pages} {229} (\bibinfo {year} {2010})}\BibitemShut
  {NoStop}%
\bibitem [{\citenamefont {Kreuzer}\ and\ \citenamefont
  {Hammer}(2011)}]{Kreuzer:2010ti}%
  \BibitemOpen
  \bibfield  {author} {\bibinfo {author} {\bibfnamefont {S.}~\bibnamefont
  {Kreuzer}}\ and\ \bibinfo {author} {\bibfnamefont {H.~W.}\ \bibnamefont
  {Hammer}},\ }\href {\doibase 10.1016/j.physletb.2010.10.003} {\bibfield
  {journal} {\bibinfo  {journal} {Phys. Lett. B}\ }\textbf {\bibinfo {volume}
  {694}},\ \bibinfo {pages} {424} (\bibinfo {year} {2011})}\BibitemShut
  {NoStop}%
\bibitem [{\citenamefont {Kreuzer}\ and\ \citenamefont
  {Grie\ss{}hammer}(2012)}]{Kreuzer:2012sr}%
  \BibitemOpen
  \bibfield  {author} {\bibinfo {author} {\bibfnamefont {S.}~\bibnamefont
  {Kreuzer}}\ and\ \bibinfo {author} {\bibfnamefont {H.~W.}\ \bibnamefont
  {Grie\ss{}hammer}},\ }\href {\doibase 10.1140/epja/i2012-12093-6} {\bibfield
  {journal} {\bibinfo  {journal} {Eur. Phys. J. A}\ }\textbf {\bibinfo {volume}
  {48}},\ \bibinfo {pages} {93} (\bibinfo {year} {2012})}\BibitemShut {NoStop}%
\bibitem [{\citenamefont {Brice\~no}\ and\ \citenamefont
  {Davoudi}(2013)}]{Briceno:2012rv}%
  \BibitemOpen
  \bibfield  {author} {\bibinfo {author} {\bibfnamefont {R.~A.}\ \bibnamefont
  {Brice\~no}}\ and\ \bibinfo {author} {\bibfnamefont {Z.}~\bibnamefont
  {Davoudi}},\ }\href {\doibase 10.1103/PhysRevD.87.094507} {\bibfield
  {journal} {\bibinfo  {journal} {Phys. Rev.}\ }\textbf {\bibinfo {volume}
  {D87}},\ \bibinfo {pages} {094507} (\bibinfo {year} {2013})}\BibitemShut
  {NoStop}%
\bibitem [{\citenamefont {Polejaeva}\ and\ \citenamefont
  {Rusetsky}(2012)}]{Polejaeva:2012ut}%
  \BibitemOpen
  \bibfield  {author} {\bibinfo {author} {\bibfnamefont {K.}~\bibnamefont
  {Polejaeva}}\ and\ \bibinfo {author} {\bibfnamefont {A.}~\bibnamefont
  {Rusetsky}},\ }\href {\doibase 10.1140/epja/i2012-12067-8} {\bibfield
  {journal} {\bibinfo  {journal} {Eur.\ Phys.\ J.\ A}\ }\textbf {\bibinfo
  {volume} {48}},\ \bibinfo {pages} {67} (\bibinfo {year} {2012})}\BibitemShut
  {NoStop}%
\bibitem [{\citenamefont {Jansen}\ \emph {et~al.}(2015)\citenamefont {Jansen},
  \citenamefont {Hammer},\ and\ \citenamefont {Jia}}]{Jansen:2015lha}%
  \BibitemOpen
  \bibfield  {author} {\bibinfo {author} {\bibfnamefont {M.}~\bibnamefont
  {Jansen}}, \bibinfo {author} {\bibfnamefont {H.~W.}\ \bibnamefont {Hammer}},
  \ and\ \bibinfo {author} {\bibfnamefont {Y.}~\bibnamefont {Jia}},\ }\href
  {\doibase 10.1103/PhysRevD.92.114031} {\bibfield  {journal} {\bibinfo
  {journal} {Phys. Rev. D}\ }\textbf {\bibinfo {volume} {92}},\ \bibinfo
  {pages} {114031} (\bibinfo {year} {2015})}\BibitemShut {NoStop}%
\bibitem [{\citenamefont {Hansen}\ and\ \citenamefont
  {Sharpe}(2016{\natexlab{a}})}]{Hansen:2015zta}%
  \BibitemOpen
  \bibfield  {author} {\bibinfo {author} {\bibfnamefont {M.~T.}\ \bibnamefont
  {Hansen}}\ and\ \bibinfo {author} {\bibfnamefont {S.~R.}\ \bibnamefont
  {Sharpe}},\ }\href {\doibase 10.1103/PhysRevD.93.014506} {\bibfield
  {journal} {\bibinfo  {journal} {Phys. Rev.}\ }\textbf {\bibinfo {volume}
  {D93}},\ \bibinfo {pages} {014506} (\bibinfo {year}
  {2016}{\natexlab{a}})}\BibitemShut {NoStop}%
\bibitem [{\citenamefont {Hansen}\ and\ \citenamefont
  {Sharpe}(2016{\natexlab{b}})}]{Hansen:2016fzj}%
  \BibitemOpen
  \bibfield  {author} {\bibinfo {author} {\bibfnamefont {M.~T.}\ \bibnamefont
  {Hansen}}\ and\ \bibinfo {author} {\bibfnamefont {S.~R.}\ \bibnamefont
  {Sharpe}},\ }\href {\doibase 10.1103/PhysRevD.96.039901,
  10.1103/PhysRevD.93.096006} {\bibfield  {journal} {\bibinfo  {journal} {Phys.
  Rev.}\ }\textbf {\bibinfo {volume} {D93}},\ \bibinfo {pages} {096006}
  (\bibinfo {year} {2016}{\natexlab{b}})},\ \bibinfo {note} {[Erratum: Phys.
  Rev. \textbf{D96}, 039901 (2017)]}\BibitemShut {NoStop}%
\bibitem [{\citenamefont {Guo}(2017)}]{Guo:2016fgl}%
  \BibitemOpen
  \bibfield  {author} {\bibinfo {author} {\bibfnamefont {P.}~\bibnamefont
  {Guo}},\ }\href {\doibase 10.1103/PhysRevD.95.054508} {\bibfield  {journal}
  {\bibinfo  {journal} {Phys. Rev.}\ }\textbf {\bibinfo {volume} {D95}},\
  \bibinfo {pages} {054508} (\bibinfo {year} {2017})}\BibitemShut {NoStop}%
\bibitem [{\citenamefont {Sharpe}(2017)}]{Sharpe:2017jej}%
  \BibitemOpen
  \bibfield  {author} {\bibinfo {author} {\bibfnamefont {S.~R.}\ \bibnamefont
  {Sharpe}},\ }\href {\doibase 10.1103/PhysRevD.96.054515} {\bibfield
  {journal} {\bibinfo  {journal} {Phys. Rev.}\ }\textbf {\bibinfo {volume}
  {D96}},\ \bibinfo {pages} {054515} (\bibinfo {year} {2017})}\BibitemShut
  {NoStop}%
\bibitem [{\citenamefont {Guo}\ and\ \citenamefont
  {Gasparian}(2018)}]{Guo:2017crd}%
  \BibitemOpen
  \bibfield  {author} {\bibinfo {author} {\bibfnamefont {P.}~\bibnamefont
  {Guo}}\ and\ \bibinfo {author} {\bibfnamefont {V.}~\bibnamefont
  {Gasparian}},\ }\href {\doibase 10.1103/PhysRevD.97.014504} {\bibfield
  {journal} {\bibinfo  {journal} {Phys. Rev. D}\ }\textbf {\bibinfo {volume}
  {97}},\ \bibinfo {pages} {014504} (\bibinfo {year} {2018})}\BibitemShut
  {NoStop}%
\bibitem [{\citenamefont {Guo}\ and\ \citenamefont
  {Gasparian}(2017)}]{Guo:2017ism}%
  \BibitemOpen
  \bibfield  {author} {\bibinfo {author} {\bibfnamefont {P.}~\bibnamefont
  {Guo}}\ and\ \bibinfo {author} {\bibfnamefont {V.}~\bibnamefont
  {Gasparian}},\ }\href {\doibase 10.1016/j.physletb.2017.10.009} {\bibfield
  {journal} {\bibinfo  {journal} {Phys. Lett.}\ }\textbf {\bibinfo {volume}
  {B774}},\ \bibinfo {pages} {441} (\bibinfo {year} {2017})}\BibitemShut
  {NoStop}%
\bibitem [{\citenamefont {Meng}\ \emph {et~al.}(2018)\citenamefont {Meng},
  \citenamefont {Liu}, \citenamefont {Mei\ss{}ner},\ and\ \citenamefont
  {Rusetsky}}]{Meng:2017jgx}%
  \BibitemOpen
  \bibfield  {author} {\bibinfo {author} {\bibfnamefont {Y.}~\bibnamefont
  {Meng}}, \bibinfo {author} {\bibfnamefont {C.}~\bibnamefont {Liu}}, \bibinfo
  {author} {\bibfnamefont {U.-G.}\ \bibnamefont {Mei\ss{}ner}}, \ and\ \bibinfo
  {author} {\bibfnamefont {A.}~\bibnamefont {Rusetsky}},\ }\href {\doibase
  10.1103/PhysRevD.98.014508} {\bibfield  {journal} {\bibinfo  {journal} {Phys.
  Rev. D}\ }\textbf {\bibinfo {volume} {98}},\ \bibinfo {pages} {014508}
  (\bibinfo {year} {2018})}\BibitemShut {NoStop}%
\bibitem [{\citenamefont {Brice\~no}\ \emph {et~al.}(2017)\citenamefont
  {Brice\~no}, \citenamefont {Hansen},\ and\ \citenamefont
  {Sharpe}}]{Briceno:2017tce}%
  \BibitemOpen
  \bibfield  {author} {\bibinfo {author} {\bibfnamefont {R.~A.}\ \bibnamefont
  {Brice\~no}}, \bibinfo {author} {\bibfnamefont {M.~T.}\ \bibnamefont
  {Hansen}}, \ and\ \bibinfo {author} {\bibfnamefont {S.~R.}\ \bibnamefont
  {Sharpe}},\ }\href {\doibase 10.1103/PhysRevD.95.074510} {\bibfield
  {journal} {\bibinfo  {journal} {Phys. Rev.}\ }\textbf {\bibinfo {volume}
  {D95}},\ \bibinfo {pages} {074510} (\bibinfo {year} {2017})}\BibitemShut
  {NoStop}%
\bibitem [{\citenamefont {Guo}\ \emph {et~al.}(2018)\citenamefont {Guo},
  \citenamefont {D\"{o}ring},\ and\ \citenamefont {Szczepaniak}}]{Guo:2018ibd}%
  \BibitemOpen
  \bibfield  {author} {\bibinfo {author} {\bibfnamefont {P.}~\bibnamefont
  {Guo}}, \bibinfo {author} {\bibfnamefont {M.}~\bibnamefont {D\"{o}ring}}, \
  and\ \bibinfo {author} {\bibfnamefont {A.~P.}\ \bibnamefont {Szczepaniak}},\
  }\href {\doibase 10.1103/PhysRevD.98.094502} {\bibfield  {journal} {\bibinfo
  {journal} {Phys. Rev.}\ }\textbf {\bibinfo {volume} {D98}},\ \bibinfo {pages}
  {094502} (\bibinfo {year} {2018})}\BibitemShut {NoStop}%
\bibitem [{\citenamefont {Guo}\ and\ \citenamefont
  {Morris}(2019)}]{Guo:2018xbv}%
  \BibitemOpen
  \bibfield  {author} {\bibinfo {author} {\bibfnamefont {P.}~\bibnamefont
  {Guo}}\ and\ \bibinfo {author} {\bibfnamefont {T.}~\bibnamefont {Morris}},\
  }\href {\doibase 10.1103/PhysRevD.99.014501} {\bibfield  {journal} {\bibinfo
  {journal} {Phys. Rev. D}\ }\textbf {\bibinfo {volume} {99}},\ \bibinfo
  {pages} {014501} (\bibinfo {year} {2019})}\BibitemShut {NoStop}%
\bibitem [{\citenamefont {Klos}\ \emph {et~al.}(2018)\citenamefont {Klos},
  \citenamefont {König}, \citenamefont {Hammer}, \citenamefont {Lynn},\ and\
  \citenamefont {Schwenk}}]{Klos:2018sen}%
  \BibitemOpen
  \bibfield  {author} {\bibinfo {author} {\bibfnamefont {P.}~\bibnamefont
  {Klos}}, \bibinfo {author} {\bibfnamefont {S.}~\bibnamefont {König}},
  \bibinfo {author} {\bibfnamefont {H.~W.}\ \bibnamefont {Hammer}}, \bibinfo
  {author} {\bibfnamefont {J.~E.}\ \bibnamefont {Lynn}}, \ and\ \bibinfo
  {author} {\bibfnamefont {A.}~\bibnamefont {Schwenk}},\ }\href {\doibase
  10.1103/PhysRevC.98.034004} {\bibfield  {journal} {\bibinfo  {journal} {Phys.
  Rev.}\ }\textbf {\bibinfo {volume} {C98}},\ \bibinfo {pages} {034004}
  (\bibinfo {year} {2018})}\BibitemShut {NoStop}%
\bibitem [{\citenamefont {Brice\~no}\ \emph {et~al.}(2018)\citenamefont
  {Brice\~no}, \citenamefont {Hansen},\ and\ \citenamefont
  {Sharpe}}]{Briceno:2018mlh}%
  \BibitemOpen
  \bibfield  {author} {\bibinfo {author} {\bibfnamefont {R.~A.}\ \bibnamefont
  {Brice\~no}}, \bibinfo {author} {\bibfnamefont {M.~T.}\ \bibnamefont
  {Hansen}}, \ and\ \bibinfo {author} {\bibfnamefont {S.~R.}\ \bibnamefont
  {Sharpe}},\ }\href {\doibase 10.1103/PhysRevD.98.014506} {\bibfield
  {journal} {\bibinfo  {journal} {Phys. Rev.}\ }\textbf {\bibinfo {volume}
  {D98}},\ \bibinfo {pages} {014506} (\bibinfo {year} {2018})}\BibitemShut
  {NoStop}%
\bibitem [{\citenamefont {Brice\~no}\ \emph
  {et~al.}(2019{\natexlab{a}})\citenamefont {Brice\~no}, \citenamefont
  {Hansen},\ and\ \citenamefont {Sharpe}}]{Briceno:2018aml}%
  \BibitemOpen
  \bibfield  {author} {\bibinfo {author} {\bibfnamefont {R.~A.}\ \bibnamefont
  {Brice\~no}}, \bibinfo {author} {\bibfnamefont {M.~T.}\ \bibnamefont
  {Hansen}}, \ and\ \bibinfo {author} {\bibfnamefont {S.~R.}\ \bibnamefont
  {Sharpe}},\ }\href {\doibase 10.1103/PhysRevD.99.014516} {\bibfield
  {journal} {\bibinfo  {journal} {Phys. Rev.}\ }\textbf {\bibinfo {volume}
  {D99}},\ \bibinfo {pages} {014516} (\bibinfo {year}
  {2019}{\natexlab{a}})}\BibitemShut {NoStop}%
\bibitem [{\citenamefont {Mai}\ \emph {et~al.}(2020)\citenamefont {Mai},
  \citenamefont {D\"{o}ring}, \citenamefont {Culver},\ and\ \citenamefont
  {Alexandru}}]{Mai:2019fba}%
  \BibitemOpen
  \bibfield  {author} {\bibinfo {author} {\bibfnamefont {M.}~\bibnamefont
  {Mai}}, \bibinfo {author} {\bibfnamefont {M.}~\bibnamefont {D\"{o}ring}},
  \bibinfo {author} {\bibfnamefont {C.}~\bibnamefont {Culver}}, \ and\ \bibinfo
  {author} {\bibfnamefont {A.}~\bibnamefont {Alexandru}},\ }\href {\doibase
  10.1103/PhysRevD.101.054510} {\bibfield  {journal} {\bibinfo  {journal}
  {Phys.\ Rev.\ D}\ }\textbf {\bibinfo {volume} {101}},\ \bibinfo {pages}
  {054510} (\bibinfo {year} {2020})}\BibitemShut {NoStop}%
\bibitem [{\citenamefont {Guo}\ and\ \citenamefont
  {D\"oring}(2020)}]{Guo:2019ogp}%
  \BibitemOpen
  \bibfield  {author} {\bibinfo {author} {\bibfnamefont {P.}~\bibnamefont
  {Guo}}\ and\ \bibinfo {author} {\bibfnamefont {M.}~\bibnamefont {D\"oring}},\
  }\href {\doibase 10.1103/PhysRevD.101.034501} {\bibfield  {journal} {\bibinfo
   {journal} {Phys. Rev. D}\ }\textbf {\bibinfo {volume} {101}},\ \bibinfo
  {pages} {034501} (\bibinfo {year} {2020})}\BibitemShut {NoStop}%
\bibitem [{\citenamefont {Guo}(2020)}]{Guo:2020spn}%
  \BibitemOpen
  \bibfield  {author} {\bibinfo {author} {\bibfnamefont {P.}~\bibnamefont
  {Guo}},\ }\href {\doibase 10.1103/PhysRevD.102.054514} {\bibfield  {journal}
  {\bibinfo  {journal} {Phys. Rev. D}\ }\textbf {\bibinfo {volume} {102}},\
  \bibinfo {pages} {054514} (\bibinfo {year} {2020})}\BibitemShut {NoStop}%
\bibitem [{\citenamefont {Blanton}\ \emph {et~al.}(2019)\citenamefont
  {Blanton}, \citenamefont {Romero-L\'opez},\ and\ \citenamefont
  {Sharpe}}]{Blanton:2019igq}%
  \BibitemOpen
  \bibfield  {author} {\bibinfo {author} {\bibfnamefont {T.~D.}\ \bibnamefont
  {Blanton}}, \bibinfo {author} {\bibfnamefont {F.}~\bibnamefont
  {Romero-L\'opez}}, \ and\ \bibinfo {author} {\bibfnamefont {S.~R.}\
  \bibnamefont {Sharpe}},\ }\href {\doibase 10.1007/JHEP03(2019)106} {\bibfield
   {journal} {\bibinfo  {journal} {JHEP}\ }\textbf {\bibinfo {volume} {03}},\
  \bibinfo {pages} {106} (\bibinfo {year} {2019})}\BibitemShut {NoStop}%
\bibitem [{\citenamefont {Pang}\ \emph {et~al.}(2019)\citenamefont {Pang},
  \citenamefont {Wu}, \citenamefont {Hammer}, \citenamefont {Mei{\ss}ner},\
  and\ \citenamefont {Rusetsky}}]{Pang:2019dfe}%
  \BibitemOpen
  \bibfield  {author} {\bibinfo {author} {\bibfnamefont {J.-Y.}\ \bibnamefont
  {Pang}}, \bibinfo {author} {\bibfnamefont {J.-J.}\ \bibnamefont {Wu}},
  \bibinfo {author} {\bibfnamefont {H.~W.}\ \bibnamefont {Hammer}}, \bibinfo
  {author} {\bibfnamefont {U.-G.}\ \bibnamefont {Mei{\ss}ner}}, \ and\ \bibinfo
  {author} {\bibfnamefont {A.}~\bibnamefont {Rusetsky}},\ }\href {\doibase
  10.1103/PhysRevD.99.074513} {\bibfield  {journal} {\bibinfo  {journal} {Phys.
  Rev.}\ }\textbf {\bibinfo {volume} {D99}},\ \bibinfo {pages} {074513}
  (\bibinfo {year} {2019})}\BibitemShut {NoStop}%
\bibitem [{\citenamefont {Jackura}\ \emph {et~al.}(2019)\citenamefont
  {Jackura}, \citenamefont {Dawid}, \citenamefont {Fern\'andez-Ram\'\i{}rez},
  \citenamefont {Mathieu}, \citenamefont {Mikhasenko}, \citenamefont {Pilloni},
  \citenamefont {Sharpe},\ and\ \citenamefont {Szczepaniak}}]{Jackura:2019bmu}%
  \BibitemOpen
  \bibfield  {author} {\bibinfo {author} {\bibfnamefont {A.~W.}\ \bibnamefont
  {Jackura}}, \bibinfo {author} {\bibfnamefont {S.~M.}\ \bibnamefont {Dawid}},
  \bibinfo {author} {\bibfnamefont {C.}~\bibnamefont
  {Fern\'andez-Ram\'\i{}rez}}, \bibinfo {author} {\bibfnamefont
  {V.}~\bibnamefont {Mathieu}}, \bibinfo {author} {\bibfnamefont
  {M.}~\bibnamefont {Mikhasenko}}, \bibinfo {author} {\bibfnamefont
  {A.}~\bibnamefont {Pilloni}}, \bibinfo {author} {\bibfnamefont {S.~R.}\
  \bibnamefont {Sharpe}}, \ and\ \bibinfo {author} {\bibfnamefont {A.~P.}\
  \bibnamefont {Szczepaniak}},\ }\href {\doibase 10.1103/PhysRevD.100.034508}
  {\bibfield  {journal} {\bibinfo  {journal} {Phys. Rev. D}\ }\textbf {\bibinfo
  {volume} {100}},\ \bibinfo {pages} {034508} (\bibinfo {year}
  {2019})}\BibitemShut {NoStop}%
\bibitem [{\citenamefont {Brice\~no}\ \emph
  {et~al.}(2019{\natexlab{b}})\citenamefont {Brice\~no}, \citenamefont
  {Hansen}, \citenamefont {Sharpe},\ and\ \citenamefont
  {Szczepaniak}}]{Briceno:2019muc}%
  \BibitemOpen
  \bibfield  {author} {\bibinfo {author} {\bibfnamefont {R.~A.}\ \bibnamefont
  {Brice\~no}}, \bibinfo {author} {\bibfnamefont {M.~T.}\ \bibnamefont
  {Hansen}}, \bibinfo {author} {\bibfnamefont {S.~R.}\ \bibnamefont {Sharpe}},
  \ and\ \bibinfo {author} {\bibfnamefont {A.~P.}\ \bibnamefont
  {Szczepaniak}},\ }\href {\doibase 10.1103/PhysRevD.100.054508} {\bibfield
  {journal} {\bibinfo  {journal} {Phys. Rev.}\ }\textbf {\bibinfo {volume}
  {D100}},\ \bibinfo {pages} {054508} (\bibinfo {year}
  {2019}{\natexlab{b}})}\BibitemShut {NoStop}%
\bibitem [{\citenamefont {Romero-L\'opez}\ \emph {et~al.}(2019)\citenamefont
  {Romero-L\'opez}, \citenamefont {Sharpe}, \citenamefont {Blanton},
  \citenamefont {Brice\~no},\ and\ \citenamefont
  {Hansen}}]{Romero-Lopez:2019qrt}%
  \BibitemOpen
  \bibfield  {author} {\bibinfo {author} {\bibfnamefont {F.}~\bibnamefont
  {Romero-L\'opez}}, \bibinfo {author} {\bibfnamefont {S.~R.}\ \bibnamefont
  {Sharpe}}, \bibinfo {author} {\bibfnamefont {T.~D.}\ \bibnamefont {Blanton}},
  \bibinfo {author} {\bibfnamefont {R.~A.}\ \bibnamefont {Brice\~no}}, \ and\
  \bibinfo {author} {\bibfnamefont {M.~T.}\ \bibnamefont {Hansen}},\ }\href
  {\doibase 10.1007/JHEP10(2019)007} {\bibfield  {journal} {\bibinfo  {journal}
  {JHEP}\ }\textbf {\bibinfo {volume} {10}},\ \bibinfo {pages} {007} (\bibinfo
  {year} {2019})}\BibitemShut {NoStop}%
\bibitem [{\citenamefont {K\"onig}(2020)}]{Konig:2020lzo}%
  \BibitemOpen
  \bibfield  {author} {\bibinfo {author} {\bibfnamefont {S.}~\bibnamefont
  {K\"onig}},\ }\href {\doibase 10.1007/s00601-020-01550-8} {\bibfield
  {journal} {\bibinfo  {journal} {Few Body Syst.}\ }\textbf {\bibinfo {volume}
  {61}},\ \bibinfo {pages} {20} (\bibinfo {year} {2020})}\BibitemShut {NoStop}%
\bibitem [{\citenamefont {Brett}\ \emph {et~al.}(2021)\citenamefont {Brett},
  \citenamefont {Culver}, \citenamefont {Mai}, \citenamefont {Alexandru},
  \citenamefont {D\"oring},\ and\ \citenamefont {Lee}}]{Brett:2021wyd}%
  \BibitemOpen
  \bibfield  {author} {\bibinfo {author} {\bibfnamefont {R.}~\bibnamefont
  {Brett}}, \bibinfo {author} {\bibfnamefont {C.}~\bibnamefont {Culver}},
  \bibinfo {author} {\bibfnamefont {M.}~\bibnamefont {Mai}}, \bibinfo {author}
  {\bibfnamefont {A.}~\bibnamefont {Alexandru}}, \bibinfo {author}
  {\bibfnamefont {M.}~\bibnamefont {D\"oring}}, \ and\ \bibinfo {author}
  {\bibfnamefont {F.~X.}\ \bibnamefont {Lee}},\ }\href {\doibase
  10.1103/PhysRevD.104.014501} {\bibfield  {journal} {\bibinfo  {journal}
  {Phys. Rev. D}\ }\textbf {\bibinfo {volume} {104}},\ \bibinfo {pages}
  {014501} (\bibinfo {year} {2021})}\BibitemShut {NoStop}%
\bibitem [{\citenamefont {Hansen}\ \emph {et~al.}(2020)\citenamefont {Hansen},
  \citenamefont {Romero-L\'opez},\ and\ \citenamefont
  {Sharpe}}]{Hansen:2020zhy}%
  \BibitemOpen
  \bibfield  {author} {\bibinfo {author} {\bibfnamefont {M.~T.}\ \bibnamefont
  {Hansen}}, \bibinfo {author} {\bibfnamefont {F.}~\bibnamefont
  {Romero-L\'opez}}, \ and\ \bibinfo {author} {\bibfnamefont {S.~R.}\
  \bibnamefont {Sharpe}},\ }\href {\doibase 10.1007/JHEP07(2020)047} {\bibfield
   {journal} {\bibinfo  {journal} {JHEP}\ }\textbf {\bibinfo {volume} {07}},\
  \bibinfo {pages} {047} (\bibinfo {year} {2020})}\BibitemShut {NoStop}%
\bibitem [{\citenamefont {Blanton}\ and\ \citenamefont
  {Sharpe}(2020{\natexlab{a}})}]{Blanton:2020gha}%
  \BibitemOpen
  \bibfield  {author} {\bibinfo {author} {\bibfnamefont {T.~D.}\ \bibnamefont
  {Blanton}}\ and\ \bibinfo {author} {\bibfnamefont {S.~R.}\ \bibnamefont
  {Sharpe}},\ }\href {\doibase 10.1103/PhysRevD.102.054520} {\bibfield
  {journal} {\bibinfo  {journal} {Phys. Rev. D}\ }\textbf {\bibinfo {volume}
  {102}},\ \bibinfo {pages} {054520} (\bibinfo {year}
  {2020}{\natexlab{a}})}\BibitemShut {NoStop}%
\bibitem [{\citenamefont {Blanton}\ and\ \citenamefont
  {Sharpe}(2020{\natexlab{b}})}]{Blanton:2020jnm}%
  \BibitemOpen
  \bibfield  {author} {\bibinfo {author} {\bibfnamefont {T.~D.}\ \bibnamefont
  {Blanton}}\ and\ \bibinfo {author} {\bibfnamefont {S.~R.}\ \bibnamefont
  {Sharpe}},\ }\href {\doibase 10.1103/PhysRevD.102.054515} {\bibfield
  {journal} {\bibinfo  {journal} {Phys. Rev. D}\ }\textbf {\bibinfo {volume}
  {102}},\ \bibinfo {pages} {054515} (\bibinfo {year}
  {2020}{\natexlab{b}})}\BibitemShut {NoStop}%
\bibitem [{\citenamefont {Pang}\ \emph {et~al.}(2020)\citenamefont {Pang},
  \citenamefont {Wu},\ and\ \citenamefont {Geng}}]{Pang:2020pkl}%
  \BibitemOpen
  \bibfield  {author} {\bibinfo {author} {\bibfnamefont {J.-Y.}\ \bibnamefont
  {Pang}}, \bibinfo {author} {\bibfnamefont {J.-J.}\ \bibnamefont {Wu}}, \ and\
  \bibinfo {author} {\bibfnamefont {L.-S.}\ \bibnamefont {Geng}},\ }\href
  {\doibase 10.1103/PhysRevD.102.114515} {\bibfield  {journal} {\bibinfo
  {journal} {Phys. Rev. D}\ }\textbf {\bibinfo {volume} {102}},\ \bibinfo
  {pages} {114515} (\bibinfo {year} {2020})}\BibitemShut {NoStop}%
\bibitem [{\citenamefont {Hansen}\ \emph {et~al.}(2021)\citenamefont {Hansen},
  \citenamefont {Brice\~no}, \citenamefont {Edwards}, \citenamefont {Thomas},\
  and\ \citenamefont {Wilson}}]{Hansen:2020otl}%
  \BibitemOpen
  \bibfield  {author} {\bibinfo {author} {\bibfnamefont {M.~T.}\ \bibnamefont
  {Hansen}}, \bibinfo {author} {\bibfnamefont {R.~A.}\ \bibnamefont
  {Brice\~no}}, \bibinfo {author} {\bibfnamefont {R.~G.}\ \bibnamefont
  {Edwards}}, \bibinfo {author} {\bibfnamefont {C.~E.}\ \bibnamefont {Thomas}},
  \ and\ \bibinfo {author} {\bibfnamefont {D.~J.}\ \bibnamefont {Wilson}},\
  }\href {\doibase 10.1103/PhysRevLett.126.012001} {\bibfield  {journal}
  {\bibinfo  {journal} {Phys. Rev. Lett.}\ }\textbf {\bibinfo {volume} {126}},\
  \bibinfo {pages} {012001} (\bibinfo {year} {2021})}\BibitemShut {NoStop}%
\bibitem [{\citenamefont {Romero-L\'opez}\ \emph {et~al.}(2021)\citenamefont
  {Romero-L\'opez}, \citenamefont {Rusetsky}, \citenamefont {Schlage},\ and\
  \citenamefont {Urbach}}]{Romero-Lopez:2020rdq}%
  \BibitemOpen
  \bibfield  {author} {\bibinfo {author} {\bibfnamefont {F.}~\bibnamefont
  {Romero-L\'opez}}, \bibinfo {author} {\bibfnamefont {A.}~\bibnamefont
  {Rusetsky}}, \bibinfo {author} {\bibfnamefont {N.}~\bibnamefont {Schlage}}, \
  and\ \bibinfo {author} {\bibfnamefont {C.}~\bibnamefont {Urbach}},\ }\href
  {\doibase 10.1007/JHEP02(2021)060} {\bibfield  {journal} {\bibinfo  {journal}
  {JHEP}\ }\textbf {\bibinfo {volume} {02}},\ \bibinfo {pages} {060} (\bibinfo
  {year} {2021})}\BibitemShut {NoStop}%
\bibitem [{\citenamefont {Blanton}\ and\ \citenamefont
  {Sharpe}(2021{\natexlab{a}})}]{Blanton:2020gmf}%
  \BibitemOpen
  \bibfield  {author} {\bibinfo {author} {\bibfnamefont {T.~D.}\ \bibnamefont
  {Blanton}}\ and\ \bibinfo {author} {\bibfnamefont {S.~R.}\ \bibnamefont
  {Sharpe}},\ }\href {\doibase 10.1103/PhysRevD.103.054503} {\bibfield
  {journal} {\bibinfo  {journal} {Phys. Rev. D}\ }\textbf {\bibinfo {volume}
  {103}},\ \bibinfo {pages} {054503} (\bibinfo {year}
  {2021}{\natexlab{a}})}\BibitemShut {NoStop}%
\bibitem [{\citenamefont {M\"{u}ller}\ \emph {et~al.}(2021)\citenamefont
  {M\"{u}ller}, \citenamefont {Rusetsky},\ and\ \citenamefont
  {Yu}}]{Muller:2020vtt}%
  \BibitemOpen
  \bibfield  {author} {\bibinfo {author} {\bibfnamefont {F.}~\bibnamefont
  {M\"{u}ller}}, \bibinfo {author} {\bibfnamefont {A.}~\bibnamefont
  {Rusetsky}}, \ and\ \bibinfo {author} {\bibfnamefont {T.}~\bibnamefont
  {Yu}},\ }\href {\doibase 10.1103/PhysRevD.103.054506} {\bibfield  {journal}
  {\bibinfo  {journal} {Phys. Rev. D}\ }\textbf {\bibinfo {volume} {103}},\
  \bibinfo {pages} {054506} (\bibinfo {year} {2021})}\BibitemShut {NoStop}%
\bibitem [{\citenamefont {Blanton}\ and\ \citenamefont
  {Sharpe}(2021{\natexlab{b}})}]{Blanton:2021mih}%
  \BibitemOpen
  \bibfield  {author} {\bibinfo {author} {\bibfnamefont {T.~D.}\ \bibnamefont
  {Blanton}}\ and\ \bibinfo {author} {\bibfnamefont {S.~R.}\ \bibnamefont
  {Sharpe}},\ }\href {\doibase 10.1103/PhysRevD.104.034509} {\bibfield
  {journal} {\bibinfo  {journal} {Phys. Rev. D}\ }\textbf {\bibinfo {volume}
  {104}},\ \bibinfo {pages} {034509} (\bibinfo {year}
  {2021}{\natexlab{b}})}\BibitemShut {NoStop}%
\bibitem [{\citenamefont {M\"uller}\ \emph {et~al.}(2022)\citenamefont
  {M\"uller}, \citenamefont {Pang}, \citenamefont {Rusetsky},\ and\
  \citenamefont {Wu}}]{Muller:2021uur}%
  \BibitemOpen
  \bibfield  {author} {\bibinfo {author} {\bibfnamefont {F.}~\bibnamefont
  {M\"uller}}, \bibinfo {author} {\bibfnamefont {J.-Y.}\ \bibnamefont {Pang}},
  \bibinfo {author} {\bibfnamefont {A.}~\bibnamefont {Rusetsky}}, \ and\
  \bibinfo {author} {\bibfnamefont {J.-J.}\ \bibnamefont {Wu}},\ }\href
  {\doibase 10.1007/JHEP02(2022)158} {\bibfield  {journal} {\bibinfo  {journal}
  {JHEP}\ }\textbf {\bibinfo {volume} {02}},\ \bibinfo {pages} {158} (\bibinfo
  {year} {2022})}\BibitemShut {NoStop}%
\bibitem [{\citenamefont {Beane}\ \emph {et~al.}(2008)\citenamefont {Beane},
  \citenamefont {Detmold}, \citenamefont {Luu}, \citenamefont {Orginos},
  \citenamefont {Savage},\ and\ \citenamefont {Torok}}]{Beane:2007es}%
  \BibitemOpen
  \bibfield  {author} {\bibinfo {author} {\bibfnamefont {S.~R.}\ \bibnamefont
  {Beane}}, \bibinfo {author} {\bibfnamefont {W.}~\bibnamefont {Detmold}},
  \bibinfo {author} {\bibfnamefont {T.~C.}\ \bibnamefont {Luu}}, \bibinfo
  {author} {\bibfnamefont {K.}~\bibnamefont {Orginos}}, \bibinfo {author}
  {\bibfnamefont {M.~J.}\ \bibnamefont {Savage}}, \ and\ \bibinfo {author}
  {\bibfnamefont {A.}~\bibnamefont {Torok}},\ }\href {\doibase
  10.1103/PhysRevLett.100.082004} {\bibfield  {journal} {\bibinfo  {journal}
  {Phys. Rev. Lett.}\ }\textbf {\bibinfo {volume} {100}},\ \bibinfo {pages}
  {082004} (\bibinfo {year} {2008})}\BibitemShut {NoStop}%
\bibitem [{\citenamefont {Detmold}\ \emph
  {et~al.}(2008{\natexlab{a}})\citenamefont {Detmold}, \citenamefont {Savage},
  \citenamefont {Torok}, \citenamefont {Beane}, \citenamefont {Luu},
  \citenamefont {Orginos},\ and\ \citenamefont {Parreno}}]{Detmold:2008fn}%
  \BibitemOpen
  \bibfield  {author} {\bibinfo {author} {\bibfnamefont {W.}~\bibnamefont
  {Detmold}}, \bibinfo {author} {\bibfnamefont {M.~J.}\ \bibnamefont {Savage}},
  \bibinfo {author} {\bibfnamefont {A.}~\bibnamefont {Torok}}, \bibinfo
  {author} {\bibfnamefont {S.~R.}\ \bibnamefont {Beane}}, \bibinfo {author}
  {\bibfnamefont {T.~C.}\ \bibnamefont {Luu}}, \bibinfo {author} {\bibfnamefont
  {K.}~\bibnamefont {Orginos}}, \ and\ \bibinfo {author} {\bibfnamefont
  {A.}~\bibnamefont {Parreno}},\ }\href {\doibase 10.1103/PhysRevD.78.014507}
  {\bibfield  {journal} {\bibinfo  {journal} {Phys. Rev.}\ }\textbf {\bibinfo
  {volume} {D78}},\ \bibinfo {pages} {014507} (\bibinfo {year}
  {2008}{\natexlab{a}})}\BibitemShut {NoStop}%
\bibitem [{\citenamefont {Detmold}\ \emph
  {et~al.}(2008{\natexlab{b}})\citenamefont {Detmold}, \citenamefont {Orginos},
  \citenamefont {Savage},\ and\ \citenamefont {Walker-Loud}}]{Detmold:2008yn}%
  \BibitemOpen
  \bibfield  {author} {\bibinfo {author} {\bibfnamefont {W.}~\bibnamefont
  {Detmold}}, \bibinfo {author} {\bibfnamefont {K.}~\bibnamefont {Orginos}},
  \bibinfo {author} {\bibfnamefont {M.~J.}\ \bibnamefont {Savage}}, \ and\
  \bibinfo {author} {\bibfnamefont {A.}~\bibnamefont {Walker-Loud}},\ }\href
  {\doibase 10.1103/PhysRevD.78.054514} {\bibfield  {journal} {\bibinfo
  {journal} {Phys. Rev. D}\ }\textbf {\bibinfo {volume} {78}},\ \bibinfo
  {pages} {054514} (\bibinfo {year} {2008}{\natexlab{b}})}\BibitemShut
  {NoStop}%
\bibitem [{\citenamefont {Blanton}\ \emph {et~al.}(2020)\citenamefont
  {Blanton}, \citenamefont {Romero-L\'opez},\ and\ \citenamefont
  {Sharpe}}]{Blanton:2019vdk}%
  \BibitemOpen
  \bibfield  {author} {\bibinfo {author} {\bibfnamefont {T.~D.}\ \bibnamefont
  {Blanton}}, \bibinfo {author} {\bibfnamefont {F.}~\bibnamefont
  {Romero-L\'opez}}, \ and\ \bibinfo {author} {\bibfnamefont {S.~R.}\
  \bibnamefont {Sharpe}},\ }\href {\doibase 10.1103/PhysRevLett.124.032001}
  {\bibfield  {journal} {\bibinfo  {journal} {Phys. Rev. Lett.}\ }\textbf
  {\bibinfo {volume} {124}},\ \bibinfo {pages} {032001} (\bibinfo {year}
  {2020})}\BibitemShut {NoStop}%
\bibitem [{\citenamefont {H\"{o}rz}\ and\ \citenamefont
  {Hanlon}(2019)}]{Horz:2019rrn}%
  \BibitemOpen
  \bibfield  {author} {\bibinfo {author} {\bibfnamefont {B.}~\bibnamefont
  {H\"{o}rz}}\ and\ \bibinfo {author} {\bibfnamefont {A.}~\bibnamefont
  {Hanlon}},\ }\href {\doibase 10.1103/PhysRevLett.123.142002} {\bibfield
  {journal} {\bibinfo  {journal} {Phys. Rev. Lett.}\ }\textbf {\bibinfo
  {volume} {123}},\ \bibinfo {pages} {142002} (\bibinfo {year}
  {2019})}\BibitemShut {NoStop}%
\bibitem [{\citenamefont {Culver}\ \emph {et~al.}(2020)\citenamefont {Culver},
  \citenamefont {Mai}, \citenamefont {Brett}, \citenamefont {Alexandru},\ and\
  \citenamefont {D\"{o}ring}}]{Culver:2019vvu}%
  \BibitemOpen
  \bibfield  {author} {\bibinfo {author} {\bibfnamefont {C.}~\bibnamefont
  {Culver}}, \bibinfo {author} {\bibfnamefont {M.}~\bibnamefont {Mai}},
  \bibinfo {author} {\bibfnamefont {R.}~\bibnamefont {Brett}}, \bibinfo
  {author} {\bibfnamefont {A.}~\bibnamefont {Alexandru}}, \ and\ \bibinfo
  {author} {\bibfnamefont {M.}~\bibnamefont {D\"{o}ring}},\ }\href {\doibase
  10.1103/PhysRevD.101.114507} {\bibfield  {journal} {\bibinfo  {journal}
  {Phys. Rev. D}\ }\textbf {\bibinfo {volume} {101}},\ \bibinfo {pages}
  {114507} (\bibinfo {year} {2020})}\BibitemShut {NoStop}%
\bibitem [{\citenamefont {Fischer}\ \emph {et~al.}(2021)\citenamefont
  {Fischer}, \citenamefont {Kostrzewa}, \citenamefont {Liu}, \citenamefont
  {Romero-L\'opez}, \citenamefont {Ueding},\ and\ \citenamefont
  {Urbach}}]{Fischer:2020jzp}%
  \BibitemOpen
  \bibfield  {author} {\bibinfo {author} {\bibfnamefont {M.}~\bibnamefont
  {Fischer}}, \bibinfo {author} {\bibfnamefont {B.}~\bibnamefont {Kostrzewa}},
  \bibinfo {author} {\bibfnamefont {L.}~\bibnamefont {Liu}}, \bibinfo {author}
  {\bibfnamefont {F.}~\bibnamefont {Romero-L\'opez}}, \bibinfo {author}
  {\bibfnamefont {M.}~\bibnamefont {Ueding}}, \ and\ \bibinfo {author}
  {\bibfnamefont {C.}~\bibnamefont {Urbach}},\ }\href {\doibase
  10.1140/epjc/s10052-021-09206-5} {\bibfield  {journal} {\bibinfo  {journal}
  {Eur. Phys. J. C}\ }\textbf {\bibinfo {volume} {81}},\ \bibinfo {pages} {436}
  (\bibinfo {year} {2021})}\BibitemShut {NoStop}%
\bibitem [{\citenamefont {Alexandru}\ \emph {et~al.}(2020)\citenamefont
  {Alexandru}, \citenamefont {Brett}, \citenamefont {Culver}, \citenamefont
  {D\"{o}ring}, \citenamefont {Guo}, \citenamefont {Lee},\ and\ \citenamefont
  {Mai}}]{Alexandru:2020xqf}%
  \BibitemOpen
  \bibfield  {author} {\bibinfo {author} {\bibfnamefont {A.}~\bibnamefont
  {Alexandru}}, \bibinfo {author} {\bibfnamefont {R.}~\bibnamefont {Brett}},
  \bibinfo {author} {\bibfnamefont {C.}~\bibnamefont {Culver}}, \bibinfo
  {author} {\bibfnamefont {M.}~\bibnamefont {D\"{o}ring}}, \bibinfo {author}
  {\bibfnamefont {D.}~\bibnamefont {Guo}}, \bibinfo {author} {\bibfnamefont
  {F.~X.}\ \bibnamefont {Lee}}, \ and\ \bibinfo {author} {\bibfnamefont
  {M.}~\bibnamefont {Mai}},\ }\href {\doibase 10.1103/PhysRevD.102.114523}
  {\bibfield  {journal} {\bibinfo  {journal} {Phys. Rev. D}\ }\textbf {\bibinfo
  {volume} {102}},\ \bibinfo {pages} {114523} (\bibinfo {year}
  {2020})}\BibitemShut {NoStop}%
\bibitem [{\citenamefont {Romero-L\'opez}\ \emph {et~al.}(2018)\citenamefont
  {Romero-L\'opez}, \citenamefont {Rusetsky},\ and\ \citenamefont
  {Urbach}}]{Romero-Lopez:2018rcb}%
  \BibitemOpen
  \bibfield  {author} {\bibinfo {author} {\bibfnamefont {F.}~\bibnamefont
  {Romero-L\'opez}}, \bibinfo {author} {\bibfnamefont {A.}~\bibnamefont
  {Rusetsky}}, \ and\ \bibinfo {author} {\bibfnamefont {C.}~\bibnamefont
  {Urbach}},\ }\href {\doibase 10.1140/epjc/s10052-018-6325-8} {\bibfield
  {journal} {\bibinfo  {journal} {Eur. Phys. J.}\ }\textbf {\bibinfo {volume}
  {C78}},\ \bibinfo {pages} {846} (\bibinfo {year} {2018})}\BibitemShut
  {NoStop}%
\bibitem [{\citenamefont {Blanton}\ \emph {et~al.}(2021)\citenamefont
  {Blanton}, \citenamefont {Hanlon}, \citenamefont {H\"orz}, \citenamefont
  {Morningstar}, \citenamefont {Romero-L\'opez},\ and\ \citenamefont
  {Sharpe}}]{Blanton:2021llb}%
  \BibitemOpen
  \bibfield  {author} {\bibinfo {author} {\bibfnamefont {T.~D.}\ \bibnamefont
  {Blanton}}, \bibinfo {author} {\bibfnamefont {A.~D.}\ \bibnamefont {Hanlon}},
  \bibinfo {author} {\bibfnamefont {B.}~\bibnamefont {H\"orz}}, \bibinfo
  {author} {\bibfnamefont {C.}~\bibnamefont {Morningstar}}, \bibinfo {author}
  {\bibfnamefont {F.}~\bibnamefont {Romero-L\'opez}}, \ and\ \bibinfo {author}
  {\bibfnamefont {S.~R.}\ \bibnamefont {Sharpe}},\ }\href {\doibase
  10.1007/JHEP10(2021)023} {\bibfield  {journal} {\bibinfo  {journal} {JHEP}\
  }\textbf {\bibinfo {volume} {10}},\ \bibinfo {pages} {023} (\bibinfo {year}
  {2021})}\BibitemShut {NoStop}%
\bibitem [{\citenamefont {Mai}\ \emph {et~al.}(2021{\natexlab{a}})\citenamefont
  {Mai}, \citenamefont {Alexandru}, \citenamefont {Brett}, \citenamefont
  {Culver}, \citenamefont {D\"oring}, \citenamefont {Lee},\ and\ \citenamefont
  {Sadasivan}}]{Mai:2021nul}%
  \BibitemOpen
  \bibfield  {author} {\bibinfo {author} {\bibfnamefont {M.}~\bibnamefont
  {Mai}}, \bibinfo {author} {\bibfnamefont {A.}~\bibnamefont {Alexandru}},
  \bibinfo {author} {\bibfnamefont {R.}~\bibnamefont {Brett}}, \bibinfo
  {author} {\bibfnamefont {C.}~\bibnamefont {Culver}}, \bibinfo {author}
  {\bibfnamefont {M.}~\bibnamefont {D\"oring}}, \bibinfo {author}
  {\bibfnamefont {F.~X.}\ \bibnamefont {Lee}}, \ and\ \bibinfo {author}
  {\bibfnamefont {D.}~\bibnamefont {Sadasivan}},\ }\href {\doibase
  10.1103/PhysRevLett.127.222001} {\bibfield  {journal} {\bibinfo  {journal}
  {Phys. Rev. Lett.}\ }\textbf {\bibinfo {volume} {127}},\ \bibinfo {pages}
  {222001} (\bibinfo {year} {2021}{\natexlab{a}})}\BibitemShut {NoStop}%
\bibitem [{\citenamefont {M\"uller}\ \emph {et~al.}(2023)\citenamefont
  {M\"uller}, \citenamefont {Pang}, \citenamefont {Rusetsky},\ and\
  \citenamefont {Wu}}]{Muller:2022oyw}%
  \BibitemOpen
  \bibfield  {author} {\bibinfo {author} {\bibfnamefont {F.}~\bibnamefont
  {M\"uller}}, \bibinfo {author} {\bibfnamefont {J.-Y.}\ \bibnamefont {Pang}},
  \bibinfo {author} {\bibfnamefont {A.}~\bibnamefont {Rusetsky}}, \ and\
  \bibinfo {author} {\bibfnamefont {J.-J.}\ \bibnamefont {Wu}},\ }\href
  {\doibase 10.1007/JHEP02(2023)214} {\bibfield  {journal} {\bibinfo  {journal}
  {JHEP}\ }\textbf {\bibinfo {volume} {02}},\ \bibinfo {pages} {214} (\bibinfo
  {year} {2023})},\ \Eprint {http://arxiv.org/abs/2211.10126} {arXiv:2211.10126
  [hep-lat]} \BibitemShut {NoStop}%
\bibitem [{\citenamefont {Blanton}\ \emph {et~al.}(2022)\citenamefont
  {Blanton}, \citenamefont {Romero-L\'opez},\ and\ \citenamefont
  {Sharpe}}]{Blanton:2021eyf}%
  \BibitemOpen
  \bibfield  {author} {\bibinfo {author} {\bibfnamefont {T.~D.}\ \bibnamefont
  {Blanton}}, \bibinfo {author} {\bibfnamefont {F.}~\bibnamefont
  {Romero-L\'opez}}, \ and\ \bibinfo {author} {\bibfnamefont {S.~R.}\
  \bibnamefont {Sharpe}},\ }\href {\doibase 10.1007/JHEP02(2022)098} {\bibfield
   {journal} {\bibinfo  {journal} {JHEP}\ }\textbf {\bibinfo {volume} {02}},\
  \bibinfo {pages} {098} (\bibinfo {year} {2022})}\BibitemShut {NoStop}%
\bibitem [{\citenamefont {Severt}\ \emph {et~al.}(2022)\citenamefont {Severt},
  \citenamefont {Mai},\ and\ \citenamefont {Mei\ss{}ner}}]{Severt:2022jtg}%
  \BibitemOpen
  \bibfield  {author} {\bibinfo {author} {\bibfnamefont {D.}~\bibnamefont
  {Severt}}, \bibinfo {author} {\bibfnamefont {M.}~\bibnamefont {Mai}}, \ and\
  \bibinfo {author} {\bibfnamefont {U.-G.}\ \bibnamefont {Mei\ss{}ner}},\
  }\href@noop {} {\  (\bibinfo {year} {2022})}\BibitemShut {NoStop}%
\bibitem [{\citenamefont {Baeza-Ballesteros}\ \emph {et~al.}(2023)\citenamefont
  {Baeza-Ballesteros}, \citenamefont {Bijnens}, \citenamefont {Husek},
  \citenamefont {Romero-L\'opez}, \citenamefont {Sharpe},\ and\ \citenamefont
  {Sj\"o}}]{Baeza-Ballesteros:2023ljl}%
  \BibitemOpen
  \bibfield  {author} {\bibinfo {author} {\bibfnamefont {J.}~\bibnamefont
  {Baeza-Ballesteros}}, \bibinfo {author} {\bibfnamefont {J.}~\bibnamefont
  {Bijnens}}, \bibinfo {author} {\bibfnamefont {T.}~\bibnamefont {Husek}},
  \bibinfo {author} {\bibfnamefont {F.}~\bibnamefont {Romero-L\'opez}},
  \bibinfo {author} {\bibfnamefont {S.~R.}\ \bibnamefont {Sharpe}}, \ and\
  \bibinfo {author} {\bibfnamefont {M.}~\bibnamefont {Sj\"o}},\ }\href@noop {}
  {\  (\bibinfo {year} {2023})},\ \Eprint {http://arxiv.org/abs/2303.13206}
  {arXiv:2303.13206 [hep-ph]} \BibitemShut {NoStop}%
\bibitem [{\citenamefont {Draper}\ \emph {et~al.}(2023)\citenamefont {Draper},
  \citenamefont {Hansen}, \citenamefont {Romero-L\'opez},\ and\ \citenamefont
  {Sharpe}}]{Draper:2023xvu}%
  \BibitemOpen
  \bibfield  {author} {\bibinfo {author} {\bibfnamefont {Z.~T.}\ \bibnamefont
  {Draper}}, \bibinfo {author} {\bibfnamefont {M.~T.}\ \bibnamefont {Hansen}},
  \bibinfo {author} {\bibfnamefont {F.}~\bibnamefont {Romero-L\'opez}}, \ and\
  \bibinfo {author} {\bibfnamefont {S.~R.}\ \bibnamefont {Sharpe}},\
  }\href@noop {} {\  (\bibinfo {year} {2023})},\ \Eprint
  {http://arxiv.org/abs/2303.10219} {arXiv:2303.10219 [hep-lat]} \BibitemShut
  {NoStop}%
\bibitem [{\citenamefont {Hansen}\ and\ \citenamefont
  {Sharpe}(2019)}]{Hansen:2019nir}%
  \BibitemOpen
  \bibfield  {author} {\bibinfo {author} {\bibfnamefont {M.~T.}\ \bibnamefont
  {Hansen}}\ and\ \bibinfo {author} {\bibfnamefont {S.~R.}\ \bibnamefont
  {Sharpe}},\ }\href {\doibase 10.1146/annurev-nucl-101918-023723} {\bibfield
  {journal} {\bibinfo  {journal} {Ann. Rev. Nucl. Part. Sci.}\ }\textbf
  {\bibinfo {volume} {69}},\ \bibinfo {pages} {65} (\bibinfo {year}
  {2019})}\BibitemShut {NoStop}%
\bibitem [{\citenamefont {Mai}\ \emph {et~al.}(2021{\natexlab{b}})\citenamefont
  {Mai}, \citenamefont {D\"oring},\ and\ \citenamefont
  {Rusetsky}}]{Mai:2021lwb}%
  \BibitemOpen
  \bibfield  {author} {\bibinfo {author} {\bibfnamefont {M.}~\bibnamefont
  {Mai}}, \bibinfo {author} {\bibfnamefont {M.}~\bibnamefont {D\"oring}}, \
  and\ \bibinfo {author} {\bibfnamefont {A.}~\bibnamefont {Rusetsky}},\ }\href
  {\doibase 10.1140/epjs/s11734-021-00146-5} {\bibfield  {journal} {\bibinfo
  {journal} {Eur. Phys. J. ST}\ }\textbf {\bibinfo {volume} {230}},\ \bibinfo
  {pages} {1623} (\bibinfo {year} {2021}{\natexlab{b}})}\BibitemShut {NoStop}%
\bibitem [{\citenamefont {Huang}\ and\ \citenamefont
  {Yang}(1957)}]{Huang:1957im}%
  \BibitemOpen
  \bibfield  {author} {\bibinfo {author} {\bibfnamefont {K.}~\bibnamefont
  {Huang}}\ and\ \bibinfo {author} {\bibfnamefont {C.~N.}\ \bibnamefont
  {Yang}},\ }\href {\doibase 10.1103/PhysRev.105.767} {\bibfield  {journal}
  {\bibinfo  {journal} {Phys. Rev.}\ }\textbf {\bibinfo {volume} {105}},\
  \bibinfo {pages} {767} (\bibinfo {year} {1957})}\BibitemShut {NoStop}%
\bibitem [{\citenamefont {Wu}(1959)}]{Wu:1959zz}%
  \BibitemOpen
  \bibfield  {author} {\bibinfo {author} {\bibfnamefont {T.~T.}\ \bibnamefont
  {Wu}},\ }\href {\doibase 10.1103/PhysRev.115.1390} {\bibfield  {journal}
  {\bibinfo  {journal} {Phys. Rev.}\ }\textbf {\bibinfo {volume} {115}},\
  \bibinfo {pages} {1390} (\bibinfo {year} {1959})}\BibitemShut {NoStop}%
\bibitem [{\citenamefont {Tan}(2008)}]{Tan:2007bg}%
  \BibitemOpen
  \bibfield  {author} {\bibinfo {author} {\bibfnamefont {S.}~\bibnamefont
  {Tan}},\ }\href {\doibase 10.1103/PhysRevA.78.013636} {\bibfield  {journal}
  {\bibinfo  {journal} {Phys.Rev.}\ }\textbf {\bibinfo {volume} {A78}},\
  \bibinfo {pages} {013636} (\bibinfo {year} {2008})}\BibitemShut {NoStop}%
\bibitem [{\citenamefont {Beane}\ \emph {et~al.}(2007)\citenamefont {Beane},
  \citenamefont {Detmold},\ and\ \citenamefont {Savage}}]{Beane:2007qr}%
  \BibitemOpen
  \bibfield  {author} {\bibinfo {author} {\bibfnamefont {S.~R.}\ \bibnamefont
  {Beane}}, \bibinfo {author} {\bibfnamefont {W.}~\bibnamefont {Detmold}}, \
  and\ \bibinfo {author} {\bibfnamefont {M.~J.}\ \bibnamefont {Savage}},\
  }\href {\doibase 10.1103/PhysRevD.76.074507} {\bibfield  {journal} {\bibinfo
  {journal} {Phys. Rev.}\ }\textbf {\bibinfo {volume} {D76}},\ \bibinfo {pages}
  {074507} (\bibinfo {year} {2007})}\BibitemShut {NoStop}%
\bibitem [{\citenamefont {Detmold}\ and\ \citenamefont
  {Savage}(2008)}]{Detmold:2008gh}%
  \BibitemOpen
  \bibfield  {author} {\bibinfo {author} {\bibfnamefont {W.}~\bibnamefont
  {Detmold}}\ and\ \bibinfo {author} {\bibfnamefont {M.~J.}\ \bibnamefont
  {Savage}},\ }\href {\doibase 10.1103/PhysRevD.77.057502} {\bibfield
  {journal} {\bibinfo  {journal} {Phys. Rev.}\ }\textbf {\bibinfo {volume}
  {D77}},\ \bibinfo {pages} {057502} (\bibinfo {year} {2008})}\BibitemShut
  {NoStop}%
\bibitem [{\citenamefont {Epelbaum}\ \emph {et~al.}(2009)\citenamefont
  {Epelbaum}, \citenamefont {Hammer},\ and\ \citenamefont
  {Mei{\ss}ner}}]{Epelbaum:2008ga}%
  \BibitemOpen
  \bibfield  {author} {\bibinfo {author} {\bibfnamefont {E.}~\bibnamefont
  {Epelbaum}}, \bibinfo {author} {\bibfnamefont {H.-W.}\ \bibnamefont
  {Hammer}}, \ and\ \bibinfo {author} {\bibfnamefont {U.-G.}\ \bibnamefont
  {Mei{\ss}ner}},\ }\href {\doibase 10.1103/RevModPhys.81.1773} {\bibfield
  {journal} {\bibinfo  {journal} {Rev. Mod. Phys.}\ }\textbf {\bibinfo {volume}
  {81}},\ \bibinfo {pages} {1773} (\bibinfo {year} {2009})}\BibitemShut
  {NoStop}%
\bibitem [{\citenamefont {L\"ahde}\ and\ \citenamefont
  {Mei{\ss}ner}(2019)}]{Lahde:2019npb}%
  \BibitemOpen
  \bibfield  {author} {\bibinfo {author} {\bibfnamefont {T.~A.}\ \bibnamefont
  {L\"ahde}}\ and\ \bibinfo {author} {\bibfnamefont {U.-G.}\ \bibnamefont
  {Mei{\ss}ner}},\ }\href {\doibase 10.1007/978-3-030-14189-9} {\emph {\bibinfo
  {title} {{Nuclear Lattice Effective Field Theory}: {An introduction}}}},\
  Vol.\ \bibinfo {volume} {957}\ (\bibinfo  {publisher} {Springer},\ \bibinfo
  {year} {2019})\BibitemShut {NoStop}%
\bibitem [{\citenamefont {Kaplan}\ \emph {et~al.}(1996)\citenamefont {Kaplan},
  \citenamefont {Savage},\ and\ \citenamefont {Wise}}]{Kaplan:1996xu}%
  \BibitemOpen
  \bibfield  {author} {\bibinfo {author} {\bibfnamefont {D.~B.}\ \bibnamefont
  {Kaplan}}, \bibinfo {author} {\bibfnamefont {M.~J.}\ \bibnamefont {Savage}},
  \ and\ \bibinfo {author} {\bibfnamefont {M.~B.}\ \bibnamefont {Wise}},\
  }\href {\doibase 10.1016/0550-3213(96)00357-4} {\bibfield  {journal}
  {\bibinfo  {journal} {Nucl. Phys. B}\ }\textbf {\bibinfo {volume} {478}},\
  \bibinfo {pages} {629} (\bibinfo {year} {1996})}\BibitemShut {NoStop}%
\bibitem [{\citenamefont {Kaplan}\ \emph
  {et~al.}(1998{\natexlab{a}})\citenamefont {Kaplan}, \citenamefont {Savage},\
  and\ \citenamefont {Wise}}]{Kaplan:1998tg}%
  \BibitemOpen
  \bibfield  {author} {\bibinfo {author} {\bibfnamefont {D.~B.}\ \bibnamefont
  {Kaplan}}, \bibinfo {author} {\bibfnamefont {M.~J.}\ \bibnamefont {Savage}},
  \ and\ \bibinfo {author} {\bibfnamefont {M.~B.}\ \bibnamefont {Wise}},\
  }\href {\doibase 10.1016/S0370-2693(98)00210-X} {\bibfield  {journal}
  {\bibinfo  {journal} {Phys. Lett. B}\ }\textbf {\bibinfo {volume} {424}},\
  \bibinfo {pages} {390} (\bibinfo {year} {1998}{\natexlab{a}})}\BibitemShut
  {NoStop}%
\bibitem [{\citenamefont {Kaplan}\ \emph
  {et~al.}(1998{\natexlab{b}})\citenamefont {Kaplan}, \citenamefont {Savage},\
  and\ \citenamefont {Wise}}]{Kaplan:1998we}%
  \BibitemOpen
  \bibfield  {author} {\bibinfo {author} {\bibfnamefont {D.~B.}\ \bibnamefont
  {Kaplan}}, \bibinfo {author} {\bibfnamefont {M.~J.}\ \bibnamefont {Savage}},
  \ and\ \bibinfo {author} {\bibfnamefont {M.~B.}\ \bibnamefont {Wise}},\
  }\href {\doibase 10.1016/S0550-3213(98)00440-4} {\bibfield  {journal}
  {\bibinfo  {journal} {Nucl. Phys. B}\ }\textbf {\bibinfo {volume} {534}},\
  \bibinfo {pages} {329} (\bibinfo {year} {1998}{\natexlab{b}})}\BibitemShut
  {NoStop}%
\bibitem [{\citenamefont {Gegelia}(1998)}]{Gegelia:1998gn}%
  \BibitemOpen
  \bibfield  {author} {\bibinfo {author} {\bibfnamefont {J.}~\bibnamefont
  {Gegelia}},\ }\href {\doibase 10.1016/S0370-2693(98)00460-2} {\bibfield
  {journal} {\bibinfo  {journal} {Phys. Lett. B}\ }\textbf {\bibinfo {volume}
  {429}},\ \bibinfo {pages} {227} (\bibinfo {year} {1998})}\BibitemShut
  {NoStop}%
\bibitem [{\citenamefont {Chen}\ \emph {et~al.}(1999)\citenamefont {Chen},
  \citenamefont {Rupak},\ and\ \citenamefont {Savage}}]{Chen:1999tn}%
  \BibitemOpen
  \bibfield  {author} {\bibinfo {author} {\bibfnamefont {J.-W.}\ \bibnamefont
  {Chen}}, \bibinfo {author} {\bibfnamefont {G.}~\bibnamefont {Rupak}}, \ and\
  \bibinfo {author} {\bibfnamefont {M.~J.}\ \bibnamefont {Savage}},\ }\href
  {\doibase 10.1016/S0375-9474(99)00298-5} {\bibfield  {journal} {\bibinfo
  {journal} {Nucl. Phys. A}\ }\textbf {\bibinfo {volume} {653}},\ \bibinfo
  {pages} {386} (\bibinfo {year} {1999})},\ \Eprint
  {http://arxiv.org/abs/nucl-th/9902056} {arXiv:nucl-th/9902056} \BibitemShut
  {NoStop}%
\bibitem [{\citenamefont {Bedaque}\ \emph {et~al.}(2000)\citenamefont
  {Bedaque}, \citenamefont {Hammer},\ and\ \citenamefont {van
  Kolck}}]{Bedaque:1999ve}%
  \BibitemOpen
  \bibfield  {author} {\bibinfo {author} {\bibfnamefont {P.}~\bibnamefont
  {Bedaque}}, \bibinfo {author} {\bibfnamefont {H.-W.}\ \bibnamefont {Hammer}},
  \ and\ \bibinfo {author} {\bibfnamefont {U.}~\bibnamefont {van Kolck}},\
  }\href {\doibase 10.1016/s0375-9474(00)00205-0} {\bibfield  {journal}
  {\bibinfo  {journal} {Nuclear Physics A}\ }\textbf {\bibinfo {volume}
  {676}},\ \bibinfo {pages} {357} (\bibinfo {year} {2000})}\BibitemShut
  {NoStop}%
\bibitem [{\citenamefont {Beane}\ \emph {et~al.}(2005)\citenamefont {Beane},
  \citenamefont {Bedaque}, \citenamefont {Parreno},\ and\ \citenamefont
  {Savage}}]{Beane:2003yx}%
  \BibitemOpen
  \bibfield  {author} {\bibinfo {author} {\bibfnamefont {S.~R.}\ \bibnamefont
  {Beane}}, \bibinfo {author} {\bibfnamefont {P.~F.}\ \bibnamefont {Bedaque}},
  \bibinfo {author} {\bibfnamefont {A.}~\bibnamefont {Parreno}}, \ and\
  \bibinfo {author} {\bibfnamefont {M.~J.}\ \bibnamefont {Savage}},\ }\href
  {\doibase 10.1016/j.nuclphysa.2004.09.081} {\bibfield  {journal} {\bibinfo
  {journal} {Nucl. Phys.}\ }\textbf {\bibinfo {volume} {A747}},\ \bibinfo
  {pages} {55} (\bibinfo {year} {2005})}\BibitemShut {NoStop}%
\bibitem [{\citenamefont {Hackenburg}(2006)}]{Hackenburg:2006qd}%
  \BibitemOpen
  \bibfield  {author} {\bibinfo {author} {\bibfnamefont {R.~W.}\ \bibnamefont
  {Hackenburg}},\ }\href {\doibase 10.1103/PhysRevC.73.044002} {\bibfield
  {journal} {\bibinfo  {journal} {Phys. Rev. C}\ }\textbf {\bibinfo {volume}
  {73}},\ \bibinfo {pages} {044002} (\bibinfo {year} {2006})}\BibitemShut
  {NoStop}%
\end{thebibliography}%


\end{document}